\documentclass[sigconf]{acmart}

\copyrightyear{2025}
\acmYear{2025}
\setcopyright{acmlicensed}
\acmConference[SIGIR '25]{Proceedings of the 48th International ACM SIGIR Conference on Research and Development in Information Retrieval}{July 13--18, 2025}{Padua, Italy}
\acmBooktitle{Proceedings of the 48th International ACM SIGIR Conference on Research and Development in Information Retrieval (SIGIR '25), July 13--18, 2025, Padua, Italy}
\acmDOI{10.1145/3726302.3730060}
\acmISBN{979-8-4007-1592-1/2025/07}

\AtBeginDocument{%
  }

\usepackage{booktabs} %

\usepackage[english]{babel}
\usepackage{moresize}
\usepackage{amsmath}
\usepackage{algorithmic}
\usepackage{balance}
\usepackage{comment}
\usepackage{paralist}
\usepackage{bm}
\usepackage{pgfplots}
\usetikzlibrary{pgfplots.dateplot}

\usepackage{flushend}
\usepackage[english]{babel}
\usepackage[latin1]{inputenc}
\usepackage{mathrsfs}
\usepackage{graphicx}

\usepackage{amssymb}
\usepackage{amsfonts}
\usepackage{url}
\usepackage{longtable}
\usepackage{rotating}
\usepackage{multirow}
\usepackage{mathrsfs}
\usepackage{subfigure}
\usepackage{enumitem}
\usepackage[linesnumbered,algoruled,boxed,lined]{algorithm2e}
\usepackage{adjustbox}
\usepackage{hyperref}
\usepackage{pgfplots}
\usetikzlibrary{pgfplots.dateplot}
\usepackage{filecontents}
\usepackage{bbm}
\usepackage{newtxmath}
\usepackage{accents} %

\newcommand{\hide}[1]{} %

\newcommand{\ve}{\textbf{e}}
\newcommand{\vece}{\textbf{e}}

\newcommand{\setu}{\mathcal{U}}
\newcommand{\setv}{\mathcal{V}}
\newcommand{\sete}{\mathcal{E}}

\def\model{UnlearnRec}

\usepackage{color}
\usepackage{colortbl}
\usepackage{xcolor}
\usepackage{hyperref}
\usepackage{makecell}
\usepackage{threeparttable}
\usepackage{tablefootnote}

\definecolor{Gray}{gray}{0.94}

\begin{document}
\title{Pre-training for Recommendation Unlearning}

\author{Guoxuan Chen}
\affiliation{%
  \institution{The University of Hong Kong}
  \country{Hong Kong}
  }
\email{guoxchen@foxmail.com}

\author{Lianghao Xia}
\affiliation{
  \institution{The University of Hong Kong}
  \country{Hong Kong}
}
\email{aka_xia@foxmail.com}

\author{Chao Huang}
\authornote{Chao Huang is the Corresponding Author.}
\affiliation{%
  \institution{The University of Hong Kong}
  \country{Hong Kong}
  }
\email{chaohuang75@gmail.com}

\begin{abstract}

Modern recommender systems powered by Graph Neural Networks (GNNs) excel at modeling complex user-item interactions, yet increasingly face scenarios requiring selective forgetting of training data. Beyond user requests to remove specific interactions due to privacy concerns or preference changes, regulatory frameworks mandate recommender systems' ability to eliminate the influence of certain user data from models. This recommendation unlearning challenge presents unique difficulties as removing connections within interaction graphs creates ripple effects throughout the model, potentially impacting recommendations for numerous users. Traditional approaches suffer from significant drawbacks: fragmentation methods damage graph structure and diminish performance, while influence function techniques make assumptions that may not hold in complex GNNs, particularly with self-supervised or random architectures. To address these limitations, we propose a novel model-agnostic pre-training paradigm \model\ that prepares systems for efficient unlearning operations. Our Influence Encoder takes unlearning requests together with existing model parameters and directly produces updated parameters of unlearned model with little fine-tuning, avoiding complete retraining while preserving model performance characteristics. Extensive evaluation on public benchmarks demonstrates that our method delivers exceptional unlearning effectiveness while providing more than 10x speedup compared to retraining approaches. 
We release our method implementation at: \textcolor{blue}{\href{https://github.com/HKUDS/UnlearnRec}{https://github.com/HKUDS/\model}}.
\end{abstract}

\vspace{-0.1in}


\maketitle

\section{Introduction}
\label{sec:intro}
Recommender systems (RS)~\cite{zhang2019deep, gao2023survey, li2024recdiff} have emerged as a critical infrastructure in modern digital landscapes, driving user engagement and satisfaction across various platforms including e-commerce, streaming services, and social media. These systems utilize vast amounts of user data to generate personalized suggestions, leveraging learning algorithms to predict user preferences and behaviors. However, growing concerns about data privacy and legal recognition of ``the right to be forgotten''~\cite{kwak2017let} through regulations like General Data Protection Regulation~\cite{protection2018general}, the Personal Information Protection and Electronic Documents Act~\cite{act2000personal} have fundamentally challenged these systems. Specifically, there is a growing need for recommendation unlearning, which involves removing specific user interactions and preferences from trained recommender models in compliance with user requests or regulatory requirements.

Graph Neural Networks (GNNs) have advanced Collaborative Filtering, establishing themselves as the leading paradigm for modern recommendation systems~\cite{wu2020comprehensive,chen2025lightgnn,zhou2020graph,xia2023automated}. These models excel at capturing complex user-item interaction graphs, yet implementing recommendation unlearning presents significant challenges due to their interconnected architecture. When users delete interactions, the process extends beyond removing data entries; the system must recalibrate recommendation probabilities for related items while ensuring deleted information becomes completely unrecoverable from both storage and inference capabilities. Crucially, this unlearning process must preserve overall prediction accuracy while making these adjustments. The challenge intensifies in GNN-based recommender systems where embeddings form an interdependent graph, collectively refined through training iterations, creating an information web that requires careful disentanglement.

Addressing recommendation unlearning challenges demands algorithms that can update models efficiently without complete retraining -- a requirement that would be computationally prohibitive in real-world applications. Current approaches fall into two primary categories: i) Partition-based Methods (P-based methods)\cite{chen2022graph,dukler2023safe,chen2022recommendation,bourtoule2021machine} and ii) Influence Function-based Methods (IF-based methods)\cite{wu2022fast,zhang2024recommendation,wu2023gif, wu2023certified}. P-based techniques strategically divide the interaction graph into multiple shards; when unlearning requests arrive, only affected shards undergo retraining before being reintegrated for prediction. However, this partitioning inevitably disrupts the graph's natural topology, degrading recommendation precision. Additionally, requests spanning numerous shards significantly increase processing time, limiting practical application. IF-based methods, while computationally faster, employ mathematical approximations to estimate unlearning impacts. These approaches face three critical limitations: their underlying mathematical assumptions often fail in complex real-world scenarios; they struggle with Self-Supervised Learning (SSL)-based GNNs where random network structures and supervision signals generated at runtime undermine gradient estimation accuracy; and they typically impose substantially higher memory requirements. These limitations highlight the urgent need for more robust and efficient unlearning solutions that maintain recommendation quality while respecting user privacy.

To address these challenges, we introduce a novel unlearning paradigm called \model, a learnable, model-agnostic framework built around a pre-trained Influence Encoder (IE). In our \model, IE can be pre-trained in advance and then either deployed directly or quickly fine-tuned when unlearning requests emerge. The core of our solution is the Influence Encoder, a learnable module specifically designed to predict how unlearning requests will impact GNN embeddings. We develop this encoder through a two-stage process of comprehensive pre-training followed by targeted fine-tuning when needed. The key contributions of our research are as follows:

\begin{itemize}[leftmargin=*]
    \item We introduce the first pretraining-based learnable paradigm for recommendation unlearning that efficiently processes requests and shifts embedding distributions while preserving accuracy.\\\vspace{-0.12in}
    \item Our \model\ framework is model-agnostic, enabling effective unlearning in state-of-the-art recommender systems.
    \item Comprehensive experiments demonstrate \model's superior performance across different dimensions.
\end{itemize}

\section{GNN-based Recommendation Unlearning}
\label{sec:model}

\subsection{Interaction Graph for Recommendation}
GNN-based methods have been shown to be the most effective solutions for CF~\cite{chen2020revisiting, xia2023automated}. In the CF paradigm, a user set $\mathcal{U}$ ($|\mathcal{U}|=I$), an item set $\setv$ ($|\setv|=J$), and a user-item interaction matrix $\textbf{A}\in\mathbb{R}^{I\times J}$ are utilized to represent the historical interaction records, based on which the encoders derive embeddings and make predictions. For each entry $a_{i,j}$ in the interaction matrix $\textbf{A}$, $a_{i,j} = 1$ if user $u_i \in \setu$ has interacted with item $v_j \in \setv$, otherwise $a_{i,j}=0$. The interaction edge set $\mathcal{E}$ corresponds one-to-one with the adjacency matrix $\textbf{A}$, i.e., if there is $a_{i,j}=1$ in $\textbf{A}$, there exists an undirected edge $(u_i, v_j) \in \mathcal{E}$.  Nowadays, GNN has demonstrated its superiority in modelling this sort of interaction graph denoted by $\mathcal{G}=(\setu, \setv, \mathcal{E})$, in which $\setu$, $\setv$ are two kinds of sets of graph nodes and $\mathcal{E}$ is the edge set.

\subsection{SSL-based GNN Recommender}
Based on interaction graph data, GNN first initializes an embedding vector for each node (user/item in $\setu,\setv$). It then performs multiple rounds of forward propagation and fusion along the edges of the graph, resulting in the final representation of last layer for each node. The final embeddings incorporate not only the information of the nodes themselves but also that of their multi-order neighbors, thereby encapsulating the structural information of the graph.
\begin{align}
    \label{eq:light_fwd}
    &\textbf{e}_{i,l} = \sum_{(v_j, u_i)\in\mathcal{E}} \frac{1}{\sqrt{\delta_i \delta_j}}\textbf{e}_{j,l-1},~~~~~
    \textbf{e}_{j,l} = \sum_{(u_i, v_j)\in\mathcal{E}} \frac{1}{\sqrt{\delta_i \delta_j}}\textbf{e}_{i,l-1}\\    
    \label{eq:light_fwd_m}
     &\textbf{E}_{l} = \textbf{D}^{-\frac{1}{2}} \cdot \bar{\textbf{A}} \cdot  \textbf{D}^{-\frac{1}{2}} \cdot \textbf{E}_{l-1},~~~~~~~\text{user}~ u_i \in\setu, ~ \text{item}~v_j \in \setv~~~~~
\end{align}
Eq.~\ref{eq:light_fwd} shows an example of GNN, which is the most widely applied backbone architecture~\cite{he2020lightgcn} for most GNN models~\cite{chen2025lightgnn,lin2022improving,wu2021self,yu2022graph}. $\textbf{e}_{i,l}, \textbf{e}_{i,l-1}\in\mathbb{R}^d$ denote the $l$-layer and $(l-1)$-layer embedding vectors for user $u_i$, and analogously for item $v_j$. Therefore, $\ve_{i,0}, \ve_{j,0}$ represent the initial embeddings for $u_i$ and $v_j$. And $\delta_i, \delta_j$ denote the degrees of nodes $u_i, v_j$, for Lapalacian normalization. Eq.~\ref{eq:light_fwd_m} gives the corresponding matrix form where $\bar{\textbf{A}} \in \mathbb{R}^{(I+J)\times(I+J)}$ is the symmetric adjacent matrix for graph $\mathcal{G}$, derived from the interaction matrix $\textbf{A}$~\cite{wang2019neural}. $\textbf{D}$ is the diagonal degree matrix of $\bar{\textbf{A}}$, and $\textbf{E}_{l} \in \mathbb{R}^{(I+J)\times d}$ denotes the nodes' embedding matrix of layer $l$, where each row in $\textbf{E}_{l}$ is an embedding for a node, i.e., $\textbf{e}_{i,l}$ or $\textbf{e}_{j,l}$. After $L$-layers' iteration, GNN aggregates the multi-order embeddings to output the final embeddings $\bar{\ve}_i,~\bar{\ve}_j \in \mathbb{R}^{d}$, corresponding to row vectors in $\bar{\textbf{E}}$, and the user-item relation prediction score $\hat{y}_{i,j}$.
\vspace{-0.05in}
\begin{align}
    \label{eq:light_readout}\hat{y}_{i,j}=\bar{\ve}_i^\top \bar{\ve}_j,~~~~~~~\bar{\ve}_i,~\bar{\ve}_j\in\bar{\textbf{E}}, ~~~~~~~\bar{\textbf{E}} = \sum_{l=0}^{L-1}\textbf{E}_l ~~~~~~~~
\end{align}
BPR loss~\cite{rendle2012bpr} is widely employed over observed interaction pairs $(u_i, v_{j^+})\in\mathcal{E}$, and sampled negative pairs $(u_i, v_{j^-})$ to optimize GNN. 
\begin{align}
    \mathcal{L}_{bpr}=\sum_{(u_i,v_{j^+}, v_{j^-})} -\log \text{sigm} (\hat{y}_{i,j^+} - \hat{y}_{i,j^-}) \label{eq:gen_bpr}
\end{align}
where $sigm(\cdot)$ denotes the Sigmoid function. By minimizing Eq.~\ref{eq:gen_bpr}, GNN recommenders can produce excellent performance but nowadays, SSL-based enhancement modules are widely used in GNNs to further achieve state-of-the-art performance. 
\begin{gather}
    \label{eq:view_func}\textbf{E}_{\mathrm{v}1}\!\! =\!\!\mathbf{View}_1(\{\textbf{E}_l | l\! \in \! [0,L\!-\!1]\}), 
    ~\textbf{E}_{\mathrm{v}2}\!\! = \!\!\mathbf{View}_2(\{\textbf{E}_l | l\! \in \! [0,L\!-\!1]\} )\\
    \label{eq:cross_view_loss}\mathcal{L}_{contrast} = \mathbf{Cross}\textbf{-}\mathbf{view}\textbf{-}\mathbf{similarity}(\textbf{E}_{\mathrm{v}1}, ~~\textbf{E}_{\mathrm{v}2})
\end{gather}
Eq.~\ref{eq:view_func},\ref{eq:cross_view_loss} give an example of SSL loss function where the model creates two views of the embeddings (Eq.~\ref{eq:view_func}) and then optimizes the contrast loss (Eq.~\ref{eq:cross_view_loss}). Besides, there are many other kinds of designs for SSL~\cite{lin2022improving,wu2021self,yu2022graph}. Of utmost importance is that, SSL-based GNNs usually generate random or data-dependent network structures and supervision signals in $\mathbf{View}_{1|2}$ functions at runtime, making it hard to precisely estimate the gradients for the trainable embeddings by manual computation without actual training (see \ref{sec:ssl_unlearn}). To recap, we use $\mathcal{M}odel(\textbf{E}_0, \bar{\textbf{A}})$ to represent and sum up GNN RS of all kinds.

\subsection{SSL-based Recommendation Unlearning}
\label{sec:ssl_unlearn}
In real-life scenarios, there is always a demand from users to undo their interaction records in order to delete accidental inputs or mistakes, change preferences, or just protect their privacy. This corresponds to deleting edges in the graph $\mathcal{G}$. We use $ \mathcal{E}_\Delta$ to denote the set of edges to be undone and $\mathcal{E}_{r}$ to denote remaining edges,i.e., $ \mathcal{E}_{\Delta} = \mathcal{E} \setminus \mathcal{E}_{r}$. In practice, pairs in $\mathcal{E}$ not only form the adjacency matrix for GNN's forward propagation but also serve as the positive supervisory signals in the loss function( Eq.~\ref{eq:gen_bpr}). Therefore, dropping a subset of edges not only changes the graph structure, but also turns the original positive labels into negative ones, definitely resulting in the distribution shift of embeddings. To avoid retraining, unlearning methods are proposed to fulfill the requirements that:  
\begin{itemize}[leftmargin=*]
    \item It must shift embeddings so that unlearned edges are predicted as negatives while maintaining a distribution similar to retraining.\\\vspace{-0.12in}
    \item Unlearned models can still achieve good prediction performance.  \\\vspace{-0.12in}
    \item Unlearning process must be more efficient than retraining.
\end{itemize}
In recommender systems, we mainly focus on edge unlearning and when some nodes need to be unlearned, we just need to unlearn all the edges associated with them, and directly delete the nodes.

For SSL-based GNNs, it is difficult to accurately estimate the difference between the old well-trained embeddings before unlearning and the newly obtained embeddings after unlearning through retraining, just by a manually-designed end-to-end function, i.e. IF like~\cite{wu2023gif,wu2023certified}, because the complicated and randomized SSL network structures and signals will produce unforeseeable embedding distributions at runtime through iterative training. So, the distribution shift estimated by IF methods would be inaccurate.

\section{Methodology}
\label{sec:solution}
\begin{figure*}[t]
    \begin{center}
    \includegraphics[width=0.99\linewidth]{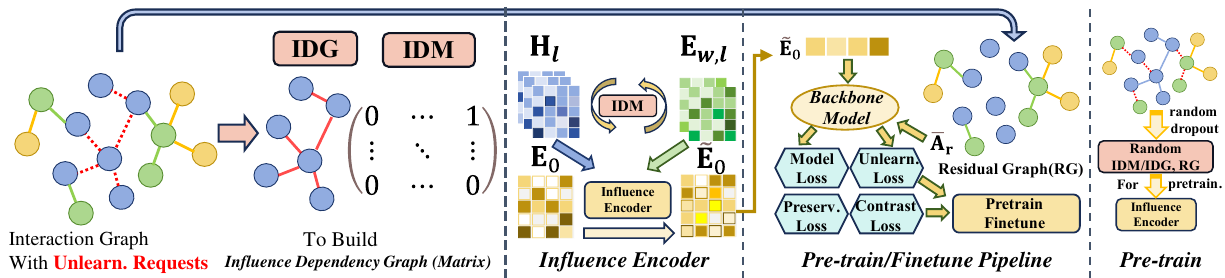}
    \end{center}
    \vspace{-0.1in}
    \caption{Overall framework of the proposed \model\ paradigm.}
    \label{fig:framework}
    \vspace{-0.1in}
\end{figure*}

Our goal is to build a model-agnostic pre-training paradigm, which produces a well-pretrained encoder $\textbf{IE}(\cdot)$ that takes the unlearning requests and the original embeddings, and outputs the unlearned embeddings directly (or for further fine-tuning, see Fig.~\ref{fig:framework}).

\subsection{Influence Dependency Matrix}
Here, we empirically derive a matrix as a means to construct the \textbf{IE}, and formalize the unlearning requests.
From the BPR loss (Eq.~\ref{eq:gen_bpr}), we can derive the approximate gradients over a positive interaction pair $(u_i,v_{j^+})$ and a negative pair $(u_i, v_{j^-})$:
\begin{align}
    \nabla_{\bar{\ve}_i} =\frac{\partial \mathcal{L}_{bpr} }{ \partial \bar{\ve}_i} &= \left(\text{sigm}(\hat{y}_{i,j^+} - \hat{y}_{i,j^-}) - 1  \right) \cdot (\bar{\ve}_{j^+} - \bar{\ve}_{j^-})\\
    \label{eq:grad}&= c \cdot (\bar{\ve}_{j^+} - \bar{\ve}_{j^-}), ~~~~\text{sigm}(\cdot)<1,~~c<0
\end{align}
These gradients will be backpropagated to every node along the graph structure during training. However, once the interaction $(u_i,v_{j^+})$ then needs to be unlearned, $\bar{\ve}_{j^+}$ should be turned into the negative side. A compensatory gradient $\nabla_{\bar{\ve}_i}^{(c)}$ should be considered:
\begin{align}
    \label{eq:grad_compen}
    \nabla^{\prime}_{\bar{\ve}_i} = \nabla_{\bar{\ve}_i} - \nabla_{\bar{\ve}_i}^{(c)}, ~~~\nabla_{\bar{\ve}_i}^{(c)}= c^{\prime} \bar{\ve}_{j^+}, ~~~c^{\prime}<c, ~|c^{\prime}|>c
\end{align}
Similarly, when a number of interaction pairs $\{(u_i,v_{j_1^+}),...,(u_i,v_{j_n^+})\}$ need to be unlearned, the compensated gradient could roughly be:
\begin{align}
    \label{eq:grad_compen_n}
    \nabla^{\prime}_{\bar{\ve}_i} = \nabla_{\bar{\ve}_i} - \nabla_{\bar{\ve}_i}^{(c)}  = \nabla_{\bar{\ve}_i} - \left( c_{i, j_1^+}^{\prime} \bar{\ve}_{j_1^+} + c_{i, j_2^+}^{\prime}\bar{\ve}_{j_2^+} +,\cdots,  c_{i, j_n^+}^{\prime}\bar{\ve}_{j_n^+}  \right)
\end{align}
The above is merely a one-order estimation. But inspired by Eq.~\ref{eq:grad}, we can find that, since $c<0$, the training process $\bar{\ve}_i^{(t+1)} = \bar{\ve}_i^{(t)} - \eta c \cdot (\bar{\ve}_{j^+}^{(t)} - \bar{\ve}_{j^-}^{(t)})$ ( learning rate $\eta>0$) is actually a process of increasing embedding similarity between observed pairs $(u_i,v_{j^+})$, and decreasing that between negative pairs $(u_i,v_{j^-})$, i.e., bringing $\bar{\ve}_i$ closer to $\bar{\ve}_{j^+}$ and away from $\bar{\ve}_{j^-}$, which aligns with the design philosophy of GNNs. Moreover, Eq.~\ref{eq:grad_compen} shows that unlearning $(u_i,v_{j^+})$ means pushing the already-pulled-closer $\bar{\ve}_{j^+}$ further away.
\begin{align}    
    \bar{\ve}_i^{(u)} &\approx \bar{\ve}_i^{(t)} - \eta \nabla_{\bar{\ve}_i^{(t)}}^{\prime} =  \bar{\ve}_i^{(t)} - \eta c \cdot (\bar{\ve}_{j^+}^{(t)} - \bar{\ve}_{j^-}^{(t)}) + \eta c^{\prime} \bar{\ve}_{j^+}^{(t)} \\
    &= \bar{\ve}_i^{(t)} - \eta(c - c^{\prime})\bar{\ve}_{j^+}^{(t)} + \eta c\bar{\ve}_{j^-}^{(t)} = \bar{\ve}_i^{(t)} - \tilde{c}_1\bar{\ve}_{j^+}^{(t)} - \tilde{c}_2\bar{\ve}_{j^-}^{(t)} 
    \label{eq:unlearn_1order}
\end{align}
Eq.~\ref{eq:unlearn_1order} is based on Eq.~\ref{eq:grad_compen}, where $\bar{\ve}_i^{(u)}$ denotes the estimated unlearned embedding, $\bar{\ve}_i^{(t)}, \bar{\ve}^{(t)}_{j^+}, \bar{\ve}_{j^-}^{(t)}$ are trained embeddings before unlearning, and $\tilde{c}_1, \tilde{c}_2 >0$. Since the original negative pair $(u_i,v_{j^-})$ has already been pushed away from each other, it is unnecessary to do that again. The only valuable part for unlearning is $\bar{\ve}_i^{(u)} \approx \bar{\ve}_i^{(t)} - \tilde{c}_1\bar{\ve}_{j^+}^{(t)}$ because only $(u_i,v_{j^+})$ is changed from positive to negative.  Extending Eq.~\ref{eq:unlearn_1order} with Eq.~\ref{eq:grad_compen_n}, we can similarly obtain:
\begin{align}
    \label{eq:unlearn_1order_n}
    \bar{\ve}_i^{(u)} \approx \bar{\ve}_i^{(t)} - \tilde{c}_{i,j_1^+}\bar{\ve}_{j_1^+}^{(t)} - \tilde{c}_{i,j_2^+}\bar{\ve}_{j_2^+}^{(t)}-,\cdots,-  \tilde{c}_{i,j_n^+}\bar{\ve}_{j_n^+}^{(t)}\\
    \label{eq:unlearn_1order_n_mat}
    \bar{\textbf{E}}^{(u)} \approx \bar{\textbf{E}}^{(t)} -  \left(\tilde{\textbf{C}} \odot \bar{\textbf{A}}_{\Delta} \right) \cdot \bar{\textbf{E}}^{(t)} ~~~~~~~~~~~~    
\end{align}
From Eq.~\ref{eq:unlearn_1order_n}, Eq.~\ref{eq:unlearn_1order_n_mat}, we find that matrix $\bar{\textbf{A}}_{\Delta}$ happens to be the  symmetric adjacent matrix constructed from the unlearning edges' set $\mathcal{E}_{\Delta}$. $\tilde{\textbf{C}}$ is the coefficient matrix and $\odot$ denotes the element-wise multiplication. Most importantly, $\bar{\textbf{A}}_{\Delta}$ actually reflects the interdependence among all unlearning requests, which collaboratively influence the direction of embedding distribution drift druing unlearning. So, we define $\bar{\textbf{A}}_{\Delta}$ as the \textbf{Influence Dependency Matrix} (IDM), the graph defined by which is correspondingly called the Influence Dependency Graph (IDG). The graph built by $\mathcal{E}_{r}$ is then referred to as Residual Graph (RG). $\bar{\textbf{A}}$ takes effect during learning, while $\bar{\textbf{A}}_{\Delta}$ takes effect during unlearning, serving as a low-pass filter.
\subsection{Learnable Influence Estimation}
\subsubsection{\bf Trainable Influence Encoder}
Inspired by IDM, we empirically propose a trainable encoder $\tilde{\textbf{E}}_0 = \textbf{IE}(\bar{\textbf{A}}_{\Delta}, \textbf{E}_0 )$ that takes the unlearning requests $\mathcal{E}_{\Delta}$ (to directly construct IDM $\bar{\textbf{A}}_{\Delta}$), and the original 0-layer embeddings of the trained model $\bm{M}$ to be unlearned as inputs, and outputs the revised 0-layer embeddings $\tilde{\textbf{E}}_0$.
\vspace{-0.05in}
\begin{align}
    \label{eq:encoder1}
     \bar{\textbf{H}} = \sum_{l=0}^{L_u-1}\textbf{H}_l, ~~~~~~\textbf{H}_{l} = \textbf{D}_{\Delta}^{-\frac{1}{2}} \cdot \bar{\textbf{A}}_{\Delta} \cdot  \textbf{D}_{\Delta}^{-\frac{1}{2}} \cdot \textbf{H}_{l-1}
\end{align}
where $\textbf{D}_{\Delta}$ is the diagonal degree matrix for $\bar{\textbf{A}}_{\Delta}$ and $\textbf{H}_{l} \in \mathbb{R}^{(I+J)\times d}$ denotes the trainable \textbf{Influence Estimation Matrix} (IEM) that will be iteratively propagated along the graph IDG for $L_u$ times, resulting in the readout IEM $\bar{\textbf{H}}$, in which each row represents a trainable influence estimation vector for each node (user/item) due to unlearning, e.g., $\textbf{h}_i,\textbf{h}_j \in \bar{\textbf{H}}$ for user $i$ and item $j$. Multiple iterations of propagation and aggregation inject more multi-hop influence dependencies and structure information of the IDG into the influence matrix IEM. Similarly, we introduce:
\vspace{-0.04in}
\begin{align}
    \label{eq:encoder2}
     \textbf{E}_{w,l} =  \textbf{D}_{\Delta}^{-\frac{1}{2}}  \bar{\textbf{A}}_{\Delta}   \textbf{D}_{\Delta}^{-\frac{1}{2}} \cdot \textbf{E}_{w,l-1},   
     ~~~\textbf{E}_{w,0} = \bar{\textbf{E}} \odot \textbf{W}_{\eta}, ~~l \in [0,L_e)    
\end{align}
where $\bar{\textbf{E}} \in \mathbb{R}^{(I+J)\times d}$ is the readout embeddings of model $\bm{M}$ before unlearning but here they are fixed and non-trainable. $\textbf{W}_{\eta} \in \mathbb{R}^{(I+J)\times 1}$ is trainable weight initialized with small values around \textbf{0}. Eq.~\ref{eq:encoder2} is inspired by Eq.~\ref{eq:unlearn_1order_n_mat} but injects higher-order information of interdependent influences. We let $\bar{\textbf{E}}_w := \textbf{E}_{w,L_e-1}$, the last layer as readout. Combining Eq.~\ref{eq:encoder1} and Eq.~\ref{eq:encoder2}, we have the final estimation:
\vspace{-0.05in}
\begin{align}
    \label{eq:encoder_out}
    \tilde{\textbf{E}}_0 =  \Delta\bar{\textbf{E}}_0 + \textbf{E}_0,  ~~\Delta\bar{\textbf{E}}_0 = \textbf{MLP}(\Delta\textbf{E}_0),  ~~~~\Delta\textbf{E}_0 = -\bar{\textbf{E}}_w +  \bar{\textbf{H}}\\
    \label{eq:encoder_mlp}
    \textbf{MLP}\left(\textbf{X}\right): ~~~~\textbf{X}_l =   \delta \left(  \textbf{W}_{l-1}\textbf{X}_{l-1} + \textbf{b}_{l-1}\right), ~~~l \in [0, L_m)~
\end{align}
where $\tilde{\textbf{E}}_0$ denotes the revised 0-layer embeddings for $\bm{M}$ after unlearning and $\textbf{MLP}(\cdot)$ denotes a multilayer perceptron in which parameters are fixed during pre-training and updated during fine-tuning. The purpose of the entire pipeline from Eq.~\ref{eq:encoder1} to Eq.~\ref{eq:encoder_out} is to  quickly calculate the embedding distribution shift $\Delta\bar{\textbf{E}}_0$ caused by unlearning when requests $\mathcal{E}_{\Delta}$ arrive. The only trainable parameters during pre-training are $\textbf{H}_0$ and $\textbf{W}_{\eta}$ that are initialized around $\textbf{0}$ and should be well pre-trained before unlearning requests come. 

\subsubsection{\bf Process Unlearning} 
We can directly construct the $\bar{\textbf{A}}_{\Delta}$ upon arrival of unlearning requests $\sete_{\Delta}$ and then calculate $\tilde{\textbf{E}}_0$ through $\textbf{IE}(\bar{\textbf{A}}_{\Delta}, \textbf{E}_0 )$. The unlearned model $\bm{M}_u$ will be $\mathcal{M}odel(\textbf{E}_0:=\tilde{\textbf{E}}_0,  ~\bar{\textbf{A}}:= \bar{\textbf{A}}_r)$ where $~\bar{\textbf{A}}_r$ is the symmetric adjacent matrix built from $\sete_r$. Parameters in $\textbf{IE}(\cdot)$ should be well-pretrained in advance.

\subsection{Multi-task Loss Functions for Pretraining}
We now introduce the loss functions for ~\model's pretraining paradigm, in which the only trainable parameters are $\textbf{H}_0$  and $\textbf{W}_{\eta}$  during pretraining. $\textbf{MLP}(\cdot)$ will be updated during fine-tuning.

\subsubsection{\bf Model Loss Function}
In Eq.~\ref{eq:model_loss},  the trained model $\bm{M}$ to be unlearned, is initialized with $\tilde{\textbf{E}}_0$ as the 0-layer embedding and computes the loss $\mathcal{L}_M$ which encompasses various complex loss functions designed by $\bm{M}$, such as SSL-based loss, BPR loss (unlearned edges should be negative) and others. $\tilde{\textbf{E}}$ is the revised final embeddings obtained from the forward and readout of $\bm{M}$ based on $\tilde{\textbf{E}}_0$ and $~\bar{\textbf{A}}_r$, which is the symmetric adjacent matrix built from $\sete_r$.
\begin{align}
    \label{eq:model_loss}
    \mathcal{L}_M = \mathcal{L}oss_{M}\left(   \mathcal{M}odel (\tilde{\textbf{E}}_0,~\bar{\textbf{A}}_r) \right), ~~~ \tilde{\textbf{E}} = \textbf{Fwd}_{M}( \tilde{\textbf{E}}_0,  ~\bar{\textbf{A}}_r ) 
\end{align}

\subsubsection{\bf Unlearning Loss}
Eq.~\ref{eq:unlearn_loss} is referred to as the unlearning loss, which is used to enforce a decrease in the predicted scores of the interaction pairs $(i_{\Delta},j_{\Delta})$ to be unlearned, as well as in the corresponding embedding similarities between $i_{\Delta}$ and $j_{\Delta}$.
\begin{align}
    \label{eq:unlearn_loss}
    \mathcal{L}_u = \sum_{(u_{i_\Delta},v_{j_\Delta})} -\log \text{sigm} ( -  {\tilde{\vece}^{\top}_{i_{\Delta}}} {\tilde{\vece}_{j_{\Delta}}} ), ~~ (i_{\Delta},j_{\Delta}) \in \mathcal{E}_{\Delta}
\end{align}
\subsubsection{\bf Preserving Loss} In Eq.~\ref{eq:emb_dist},
$\bm{\Psi}( \bar{\textbf{E}} )$ generates a vector describing  the embedding distribution of remaining positive pairs $(u_i,v_{j^+})$ in $\sete_r$ based on $\bm{M}$'s original final embeddings $\bar{\textbf{E}}$. Analogously, we can obtain $\bm{\Psi}( \tilde{\textbf{E}} )$ for the distribution of revised final embeddings $\tilde{\textbf{E}}$.
\begin{align}
    \label{eq:emb_dist}
    \bm{\Psi}( \bar{\textbf{E}} ) = \left[  \text{log} \frac{ \text{exp}  \left( \bar{\vece}_i^{\top} \bar{\vece}_{j^+} / \tau  \right) }{  \sum_{(*,v_{k^+}) \in \sete_r} \text{exp} \left(\bar{\vece}_i^{\top} \bar{\vece}_{k^+} / \tau \right) }, \cdots   \bigg|  ~~ (u_i, v_{j^+}) \in \sete_r \right]~~~~~~~
\end{align}
 We can align the two vectors because typically, the unlearning set $\sete_{\Delta}$ represents only a small proportion of the entire $\sete$. As a result, the embedding distribution of the remaining unaffected portion retains a significant amount of useful information. This is beneficial to preserving predictive performance of $\bm M$ after unlearning.
\begin{align}
    \label{eq:keep_loss}
    \mathcal{L}_p = \textbf{Align}\left( \bm{\Psi}( \tilde{\textbf{E}} ),  ~~\bm{\Psi}( \bar{\textbf{E}} )  \right)
\end{align}
Eq.~\ref{eq:keep_loss} is named preserving loss and we use $L_2$ distance as $\textbf{Align}(\cdot)$.
\subsubsection{\bf Contrast Loss}
The distinct aspect of recommending unlearning compared to general GNN unlearning lies in that when a user unlearns certain interactions, it implies, to some extent, a decrease in the probability of recommending similar interactions.
\begin{align}
    \label{eq:contr_loss_fwd}
     \textbf{H}^{\prime}_{l} = \textbf{D}_{\Delta^{\prime}}^{-\frac{1}{2}} \cdot \bar{\textbf{A}}^{\prime}_{\Delta} \cdot  \textbf{D}_{\Delta^{\prime}}^{-\frac{1}{2}} \cdot \textbf{H}_{l-1}, ~~~\bar{\textbf{A}}^{\prime}_{\Delta} = \textbf{Dropout}(\bar{\textbf{A}}_{\Delta})
\end{align}
$\textbf{Dropout}(\cdot)$ randomly removes a small portion (e.g. $\rho$\%) of unlearning edges from $\bar{\textbf{A}}_{\Delta}$ and is then used for forward propagation. Next, we align the generated $\textbf{H}^{\prime}_{l}$ by $\textbf{Dropout}(\cdot)$ with the original $\textbf{H}_{l}$.
\begin{align}
    \label{eq:contr_loss}
    \mathcal{L}_{c} =  \sum_{i\in \setu \cup\setv}  - \text{log} \frac{ \text{exp}\left( 
   \text{cos}( \textbf{h}_i, \textbf{h}^{\prime}_i ) / \tau \right)    }{ \sum_{i^{\prime} \in \setu \cup\setv }\text{exp}\left(  cos(  \textbf{h}_{i^{\prime}}, \textbf{h}^{\prime}_{i^{\prime}} ) / \tau    \right) } 
\end{align} 
where $\textbf{h}_i, \textbf{h}^{\prime}_i$ denote row vectors in $\textbf{H}_{l}$, $\textbf{H}^{\prime}_{l}$,respectively.
To some extent, it leverages the $\textbf{H}^{\prime}_{l}$ generated on $\bar{\textbf{A}}^{\prime}_{\Delta}$ to predict the complete $\textbf{H}_{l}$, thereby implicitly utilizing a partial IDM $\bar{\textbf{A}}^{\prime}_{\Delta}$ to complete and predict the full IDM $\bar{\textbf{A}}_{\Delta}$,  by which, it enables the incorporation of more contextual influence and correlations. To summarize all of the above, we we obtain the final loss function for pre-training:
\begin{align}
    \label{eq:final_loss}
    \mathcal{L} = \mathcal{L}_M + \lambda_u \mathcal{L}_u + \lambda_p \mathcal{L}_p + \lambda_c \mathcal{L}_{c} = \mathcal{L}oss\left(   
    \bm{M}, \bar{\textbf{A}}_{\Delta}    
    \right)
\end{align} 
where $\mathcal{L}_M$ contains all the SSL random network structures of model $\bm M$, which will drive the post-unlearning embeddings to conform to the specific distribution characteristics of the model. $\mathcal{L}_p$ is responsible for maintaining the predictive performance while $\mathcal{L}_u$ is used for unlearning, and they represent a trade-off. $\mathcal{L}_{c}$ controls the influence generalization. The final loss $\mathcal{L}$ can be computed as long as the IDM $\bar{\textbf{A}}_{\Delta}$ and the trained model $\bm{M}$ to be unlearned are given.

\subsection{Pretraining Paradigm for Unlearning}
We train the \textbf{IE} during the pretraining and adjust both the \textbf{IE} and the RS model based on unlearning requests during the finetuning.
\subsubsection{\bf Pre-training}
In each simulation round, we randomly select a subset $\sete_{\Delta}^{(s)}$ from $\sete$ (e.g., $\rho$ \%) as the simulated unlearning set to construct the IDM $\bar{\textbf{A}}_{\Delta}^{(s)}$. Then, we optimize the loss function $\mathcal{L}$ to train the parameters. Currently, only $\textbf{H}_0$ in Eq.~\ref{eq:encoder1} and $\textbf{W}_{\eta}$ in Eq.~\ref{eq:encoder2} are trainable, while the other parameters like $\bar{\textbf{E}}$ remain fixed. Please refer to lines 1 to 8 of Algorithm~\ref{alg:pretrain_alg} for more details.
\subsubsection{\bf Fine-tuning} The well-pretrained $\textbf{IE}(\cdot)$ can be directly utilized to perform unlearning. Howver, to achieve better performance and unlearning efficacy, if necessary, we can still fine-tune $\textbf{IE}(\cdot)$ and the unlearned model $\bm{M}_u$ when actual unlearning requests arrive. During fine-tuning, we usually only need to optimize $\mathcal{L}_M + \lambda_u \mathcal{L}_u$ to fine-tune the $\textbf{MLP}(\cdot)$ in $\textbf{IE}(\cdot)$ (Eq.~\ref{eq:encoder_mlp}) and $\textbf{E}_0$ in $\bm{M}_u$ (Eq.~\ref{eq:light_fwd}), which enables us to accomplish the fine-tuning process very efficiently. Please refer to lines 9 and 10 of Algorithm~\ref{alg:pretrain_alg} for more information.
\vspace{-0.1in}
\begin{algorithm}[h]
    \caption{Pre-train./Fine-tun. Paradigm of~\model}
    \label{alg:pretrain_alg}
    \KwIn{
        Trained model $\bm{M}$ and graph data $\mathcal{G}=(\setu, \setv, \mathcal{E})$
    }
    \KwOut{
        Well-pretrained $\textbf{IE}(\cdot)$ and unlearned $\bm{M}_u$. 
    }
    Initialize $\textbf{H}_0$ in Eq.~\ref{eq:encoder1} and $\textbf{W}_{\eta}$ in Eq.~\ref{eq:encoder2} around $\textbf{0}$. 
    Initialize $\textbf{W}_l$,$\textbf{b}_l$ in Eq.~\ref{eq:encoder_mlp} with identity matrix and $\textbf{0}$, respectively \\
    Lock all other parameters and leave only $\textbf{H}_0$,$\textbf{W}_{\eta}$ trainable\\
  
    \For{run $i_{pre}=1$ \KwTo $N_{pretrain}$}{
        Sample a subset $\sete_{\Delta}^{(s)}$ from $\sete$ to construct IDM $\bar{\textbf{A}}_{\Delta}^{(s)}$
        \\
        \For{epoch $j_{tr}=1$ \KwTo $N_{train}$}{            
            Minimize $\mathcal{L}oss(\bm{M},\bar{\textbf{A}}^{(s)}_{\Delta})$ (Eq.~\ref{eq:final_loss}) to update $\textbf{H}_0$, $\textbf{W}_{\eta}$;\\
        }
    }
    Actual unlearning requests $\sete_{\Delta}$ come, $\mathcal{E}_{r} = \mathcal{E} \setminus \mathcal{E}_{\Delta}$, build $\bar{\textbf{A}}_r$.\\ 
    Fix all parameters except those in $\textbf{MLP}(\cdot)$ and $\textbf{E}_0$. Minimize $\mathcal{L}_M + \lambda_u \mathcal{L}_u$ based on ($\tilde{\textbf{E}}_0$,$\bar{\textbf{A}}_r$) to fine-tune $\textbf{MLP}(\cdot)$ and $\textbf{E}_0$.\\
    \textbf{Return} Well-pretrained $\textbf{IE}(\cdot)$ and $\bm{M}_u = \mathcal{M}odel(\tilde{\textbf{E}}_0,   \bar{\textbf{A}}_r)$
\end{algorithm}

\vspace{-0.1in}
\section{Evaluation}
\label{sec:exp}
We conducted numerous experiments to validate our paradigm \model and address the following research questions (RQs):
\begin{itemize}[leftmargin=*]
    \item \textbf{RQ1}: Can the proposed paradigm \model\ effectively preserve the predictive performance of the model after unlearning, compared to baselines and retrained model?\\\vspace{-0.12in}
    \item \textbf{RQ2}: Can our~\model\ paradigm effectively unlearn the requested interactions in comparison to the baseline approaches?\\\vspace{-0.12in}
    \item \textbf{RQ3}: 
     How do the components of the proposed~\model\ paradigm affect the effectiveness of unlearning?
    \\\vspace{-0.12in}
    \item \textbf{RQ4}: 
    How  does our~\model\ method alter the distribution curve of the embeddings of the model that needs to be unlearned?
    \\\vspace{-0.12in}
    \item \textbf{RQ5}: How efficient is our \model\ approach in comparison to existing techniques in terms of time and memory?\\\vspace{-0.12in}
    \item \textbf{RQ6}: Can the~\model\ method effectively reduce the recommended probability of items similar to the unlearned items?
\end{itemize}

\subsection{Experimental Settings}
\subsubsection{\bf Evaluation Datasets}
\begin{table}[t]
    \centering 
    \small
    \caption{Statistical details of experimental datasets.}
    \label{tab:stat}
    \vspace{-0.1in}
    \setlength{\tabcolsep}{1.1mm}
    \begin{tabular}{ccccc}
    \hline
    Dataset &\# Users  &\# Items  &\# Interactions &Interaction Density \\
    \hline
    \hline
    Gowalla &25557  &19747  &294983 & $5.85 \times 10^{-4}$ \\
    Yelp2018    &31668  &38048  &1561406 &  $1.30 \times 10^{-3} $ \\
    Movielens-1M  &6940  &3706  &1000209  & $3.89 \times 10^{-2} $  \\
    \hline
    \end{tabular}  
    \vspace{-0.1in}
\end{table}

We validate the unlearning effectiveness of our \model\ using three commonly used real-world datasets: Movielens-1M, Gowalla, and Yelp2018. 
\textbf{Movielens-1M} is widely used public user behavior collection, containing one million ratings from users on movies while the \textbf{Gowalla} dataset comprises user check-in records at geographic locations between January and June 2010, collected from the Gowalla platform. And the \textbf{Yelp2018} dataset is derived from the Yelp platform and includes user ratings on venues spanning January to June 2018. For training the model $\bm{m}$ to be unlearned, we allocate 20\% of the total interaction edges as the test set, while the remaining edges are utilized for training.
\subsubsection{\bf Prediction Metrics}
Following the standard evaluation protocols commonly used for the CF tasks~\cite{wang2019neural, zhang2022incorporating}, We employ a full-rank evaluation approach where we rank all uninteracted items alongside the positive items from the test set for each user. We utilize two widely used metrics, namely Recall@N and NDCG@N~\cite{wang2023diffusion, wu2021self}, to assess the prediction performance of the trained model.

\subsubsection{\bf Unlearning Metrics}
Following the common validation methods of other unlearning works~\cite{wu2023gif,wu2023certified}, we conduct adversarial attacks and utilize the \textit{Membership Inference} (MI)~\cite{olatunji2021membership} as the metric to evaluate the unlearning effectiveness. MI can indicate whether the unlearned model leaks information about the unlearned edges, specifically whether the existence of unlearned edges can be inferred from the unlearned model's embeddings, or remain completely concealed. We use MI-BF to denote the ratio of the average probability for presence of edges in $\sete_{\Delta}$ before and after unlearning, and MI-NG to measure the ratio between average recommended probability of unlearned edges in model $\bm{M}_u$ and probability of edges obtained through negative sampling. MI-BF and MI-NG should exceed 1 and the higher, the better.

\subsubsection{\bf Adversarial Attacks}
Following other unlearning works~\cite{wu2023gif,wu2023certified}, we conduct adversarial attacks to validate our proposed paradigm. Specifically, we select adversarial edges by randomly sampling the least probable edges, with their probability based on a GCN trained on the original dataset. Afterwards, we train the GNN models using the attacked graph that incorporates these adversarial edges. Subsequently, we employ all baseline unlearning methods to unlearn the adversarial edges, and evaluate the unlearning performance of each baseline using the aforementioned metrics.

\subsubsection{\bf Backbone Models}
We employ the following GNN models as our backbone models and the targets for unlearning to evaluate the effectiveness of our paradigm in unlearning.
\begin{itemize}[leftmargin=*]
    \item \textbf{LightGCN}~\cite{he2020lightgcn} utilizes fundamental GCN architectures to enhance performance in recommendation tasks ("GCN" for short).\\\vspace{-0.12in}
    \item \textbf{SGL}~\cite{wu2021self} presents multiple techniques for enhancing graph contrastive learning through graph and feature augmentations \\\vspace{-0.12in}
    \item \textbf{SimGCL}~\cite{yu2022graph} is a contrastive learning model with simple feature-level augmentation techniques utilizing random permutation.\\\vspace{-0.12in}
\end{itemize}

\subsubsection{\bf Baseline Methods}
We compare our \model\ paradigm with the state-of-the-art baselines from various  perspectives.

\begin{itemize}[leftmargin=*]
    \item \textbf{GIF}~\cite{wu2023gif} is a state-of-the-art method for graph unlearning that employs the influence function (IF) to estimate the shifts in parameters related to the graph structures needing to be unlearned.\\\vspace{-0.12in}
    \item \textbf{CEU}~\cite{wu2023certified} is a state-of-the-art IF-based method designed to expedite the unlearning process and providing theoretical guarantee.\\\vspace{-0.12in}
    \item \textbf{GraphEraser}~\cite{chen2022graph} is a partition-based approach for unlearning, where the graph is divided into multiple shards based on nodes using embedding clustering and community detection techniques. These shards are subsequently combined and when doing unlearning, the method just selectively retrains the relevant shards.\\\vspace{-0.12in}
    \item \textbf{RecEraser}~\cite{chen2022recommendation} is a recent partition-based method for recommendation unlearning, which differs from GraphEraser by partitioning the edges instead of nodes into shards to avoid losing edge information. Additionally, it incorporates a more sophisticated attention aggregation method with more parameters.
    \\\vspace{-0.12in}
    \item \textbf{Retrain}. When unlearning requests are received, a full retraining of the entire graph model is carried out with model parameters being re-initialized. This can be regraded as ground truth.
\end{itemize}

\begin{table*}[t]
    \centering
    \caption{Overall unlearning utility and efficacy comparison on Gowalla, Movielens-1M, and Yelp2018 datasets.}%
    \label{tab:main_tab}
    \vspace{-0.1in}
    \footnotesize
    \setlength{\tabcolsep}{0.8mm}

\begin{tabular}{|cc|ccc|ccc|ccc|ccc|ccc|ccc|}
\hline
\multicolumn{2}{|c|}{Method}                            & \multicolumn{3}{c|}{Retrain}                                        & \multicolumn{3}{c|}{RecEraser}                                     & \multicolumn{3}{c|}{GraphEraser}                                   & \multicolumn{3}{c|}{CEU}                                           & \multicolumn{3}{c|}{GIF}                                           & \multicolumn{3}{c|}{Ours}                                                                     \\ \hline
\multicolumn{2}{|c|}{Backbone}                          & \multicolumn{1}{c|}{GCN}    & \multicolumn{1}{c|}{SimGCL}  & SGL    & \multicolumn{1}{c|}{GCN}    & \multicolumn{1}{c|}{SimGCL} & SGL    & \multicolumn{1}{c|}{GCN}    & \multicolumn{1}{c|}{SimGCL} & SGL    & \multicolumn{1}{c|}{GCN}    & \multicolumn{1}{c|}{SimGCL} & SGL    & \multicolumn{1}{c|}{GCN}    & \multicolumn{1}{c|}{SimGCL} & SGL    & \multicolumn{1}{c|}{GCN}             & \multicolumn{1}{c|}{SimGCL}          & SGL             \\\rowcolor{Gray} \Xhline{0.7px}
\multicolumn{1}{|c|}{\multirow{4}{*}{ML-1m}}   & Recall & \multicolumn{1}{c|}{0.2287} & \multicolumn{1}{c|}{0.2336}  & 0.2339 & \multicolumn{1}{c|}{0.2033} & \multicolumn{1}{c|}{0.2106} & 0.2099 & \multicolumn{1}{c|}{0.2008} & \multicolumn{1}{c|}{0.1995} & 0.1989 & \multicolumn{1}{c|}{0.2219} & \multicolumn{1}{c|}{0.2239} & 0.2234 & \multicolumn{1}{c|}{0.2111} & \multicolumn{1}{c|}{0.2117} & 0.2177 & \multicolumn{1}{c|}{0.2282}          & \multicolumn{1}{c|}{0.2336}          & 0.2330          \\  \cline{2-20} 
\multicolumn{1}{|c|}{}                         & NDCG   & \multicolumn{1}{c|}{0.3209} & \multicolumn{1}{c|}{0.3325}  & 0.3328 & \multicolumn{1}{c|}{0.2807} & \multicolumn{1}{c|}{0.2978} & 0.2887 & \multicolumn{1}{c|}{0.2709} & \multicolumn{1}{c|}{0.2752} & 0.2797 & \multicolumn{1}{c|}{0.3152} & \multicolumn{1}{c|}{0.3217} & 0.3163 & \multicolumn{1}{c|}{0.2912} & \multicolumn{1}{c|}{0.3067} & 0.3031 & \multicolumn{1}{c|}{0.3206}          & \multicolumn{1}{c|}{0.3323}          & 0.3296          \\ \cline{2-20} 
\multicolumn{1}{|c|}{}                         & MI-BF  & \multicolumn{1}{c|}{9.7865} & \multicolumn{1}{c|}{6.2314}  & 20.984 & \multicolumn{1}{c|}{1.8103} & \multicolumn{1}{c|}{1.8863} & 1.9968 & \multicolumn{1}{c|}{1.7327} & \multicolumn{1}{c|}{1.9054} & 2.1676 & \multicolumn{1}{c|}{1.0601} & \multicolumn{1}{c|}{1.8883} & 1.1281 & \multicolumn{1}{c|}{1.0426} & \multicolumn{1}{c|}{1.5980} & 1.1063 & \multicolumn{1}{c|}{6.5071}          & \multicolumn{1}{c|}{3.1981}          & 14.0285         \\ \cline{2-20} 
\multicolumn{1}{|c|}{}                         & MI-NG  & \multicolumn{1}{c|}{6.2327} & \multicolumn{1}{c|}{3.9845}  & 11.344 & \multicolumn{1}{c|}{1.6734} & \multicolumn{1}{c|}{1.2572} & 1.6589 & \multicolumn{1}{c|}{1.3160} & \multicolumn{1}{c|}{1.2202} & 1.8691 & \multicolumn{1}{c|}{0.7462} & \multicolumn{1}{c|}{1.8018} & 0.8396 & \multicolumn{1}{c|}{0.9351} & \multicolumn{1}{c|}{1.6629} & 0.8123 & \multicolumn{1}{c|}{\textbf{4.2766}} & \multicolumn{1}{c|}{\textbf{2.3575}} & \textbf{9.3661} \\\rowcolor{Gray} \Xhline{0.7px}
\multicolumn{1}{|c|}{\multirow{4}{*}{Gowalla}} & Recall & \multicolumn{1}{c|}{0.2407} & \multicolumn{1}{c|}{0.2605}  & 0.2502 & \multicolumn{1}{c|}{0.2211} & \multicolumn{1}{c|}{0.2367} & 0.2279 & \multicolumn{1}{c|}{0.2191} & \multicolumn{1}{c|}{0.2238} & 0.2213 & \multicolumn{1}{c|}{0.2390} & \multicolumn{1}{c|}{0.2516} & 0.2421 & \multicolumn{1}{c|}{0.2391} & \multicolumn{1}{c|}{0.2486} & 0.2394 & \multicolumn{1}{c|}{0.2398}          & \multicolumn{1}{c|}{0.2540}          & 0.2477          \\ \cline{2-20} 
\multicolumn{1}{|c|}{}                         & NDCG   & \multicolumn{1}{c|}{0.1584} & \multicolumn{1}{c|}{0.1703}  & 0.1618 & \multicolumn{1}{c|}{0.1432} & \multicolumn{1}{c|}{0.1453} & 0.1486 & \multicolumn{1}{c|}{0.1399} & \multicolumn{1}{c|}{0.1453} & 0.1429 & \multicolumn{1}{c|}{0.1505} & \multicolumn{1}{c|}{0.1618} & 0.1558 & \multicolumn{1}{c|}{0.1508} & \multicolumn{1}{c|}{0.1599} & 0.1524 & \multicolumn{1}{c|}{0.1545}          & \multicolumn{1}{c|}{0.1670}          & 0.1611          \\ \cline{2-20} 
\multicolumn{1}{|c|}{}                         & MI-BF  & \multicolumn{1}{c|}{7.6745} & \multicolumn{1}{c|}{15.4290} & 8.7485 & \multicolumn{1}{c|}{1.4975} & \multicolumn{1}{c|}{1.2150} & 1.7941 & \multicolumn{1}{c|}{1.5865} & \multicolumn{1}{c|}{1.0238} & 1.8937 & \multicolumn{1}{c|}{1.1470} & \multicolumn{1}{c|}{1.0322} & 1.1283 & \multicolumn{1}{c|}{1.0030} & \multicolumn{1}{c|}{1.0432} & 1.0076 & \multicolumn{1}{c|}{5.3729}          & \multicolumn{1}{c|}{4.3520}          & 2.6285          \\ \cline{2-20} 
\multicolumn{1}{|c|}{}                         & MI-NG  & \multicolumn{1}{c|}{3.9265} & \multicolumn{1}{c|}{7.7154}  & 4.4323 & \multicolumn{1}{c|}{0.9473} & \multicolumn{1}{c|}{0.8916} & 1.0357 & \multicolumn{1}{c|}{0.8182} & \multicolumn{1}{c|}{0.6511} & 0.9995 & \multicolumn{1}{c|}{0.5345} & \multicolumn{1}{c|}{0.5644} & 0.5317 & \multicolumn{1}{c|}{0.5059} & \multicolumn{1}{c|}{0.5883} & 0.5108 & \multicolumn{1}{c|}{\textbf{2.8091}} & \multicolumn{1}{c|}{\textbf{1.8075}} & \textbf{1.3369} \\  \rowcolor{Gray} \Xhline{0.7px}
\multicolumn{1}{|c|}{\multirow{4}{*}{Yelp}}    & Recall & \multicolumn{1}{c|}{0.0625} & \multicolumn{1}{c|}{0.0648}  & 0.0635 & \multicolumn{1}{c|}{0.0587} & \multicolumn{1}{c|}{0.0614} & 0.0599 & \multicolumn{1}{c|}{0.0574} & \multicolumn{1}{c|}{0.0610} & 0.0578 & \multicolumn{1}{c|}{0.0597} & \multicolumn{1}{c|}{0.0633} & 0.0621 & \multicolumn{1}{c|}{0.0580} & \multicolumn{1}{c|}{0.0630} & 0.0624 & \multicolumn{1}{c|}{0.0617}          & \multicolumn{1}{c|}{0.0640}          & 0.0633          \\ \cline{2-20} 
\multicolumn{1}{|c|}{}                         & NDCG   & \multicolumn{1}{c|}{0.0511} & \multicolumn{1}{c|}{0.0531}  & 0.0512 & \multicolumn{1}{c|}{0.0475} & \multicolumn{1}{c|}{0.0501} & 0.0482 & \multicolumn{1}{c|}{0.0466} & \multicolumn{1}{c|}{0.0478} & 0.0460 & \multicolumn{1}{c|}{0.0493} & \multicolumn{1}{c|}{0.0516} & 0.0497 & \multicolumn{1}{c|}{0.0461} & \multicolumn{1}{c|}{0.0513} & 0.0501 & \multicolumn{1}{c|}{0.0501}          & \multicolumn{1}{c|}{0.0519}          & 0.0513          \\ \cline{2-20} 
\multicolumn{1}{|c|}{}                         & MI-BF  & \multicolumn{1}{c|}{6.2634} & \multicolumn{1}{c|}{11.1054} & 9.9890 & \multicolumn{1}{c|}{1.5473} & \multicolumn{1}{c|}{1.2490} & 1.4003 & \multicolumn{1}{c|}{1.4832} & \multicolumn{1}{c|}{1.2837} & 1.3896 & \multicolumn{1}{c|}{1.1415} & \multicolumn{1}{c|}{1.4724} & 1.0751 & \multicolumn{1}{c|}{1.0113} & \multicolumn{1}{c|}{1.5555} & 1.0503 & \multicolumn{1}{c|}{5.6431}          & \multicolumn{1}{c|}{6.0197}          & 4.0866          \\ \cline{2-20} 
\multicolumn{1}{|c|}{}                         & MI-NG  & \multicolumn{1}{c|}{3.3154} & \multicolumn{1}{c|}{5.8891}  & 4.4316 & \multicolumn{1}{c|}{0.9043} & \multicolumn{1}{c|}{0.6777} & 0.6195 & \multicolumn{1}{c|}{0.8968} & \multicolumn{1}{c|}{0.6873} & 0.6536 & \multicolumn{1}{c|}{0.6055} & \multicolumn{1}{c|}{0.7977} & 0.5801 & \multicolumn{1}{c|}{0.5188} & \multicolumn{1}{c|}{0.8539} & 0.5689 & \multicolumn{1}{c|}{2.9978}          & \multicolumn{1}{c|}{2.4956}          & 2.0492          \\ \hline
\end{tabular}

\vspace{-0.1in}
\end{table*}

\subsubsection{\bf Implementation Settings}
We develop our \model\ using PyTorch and employ the Adam optimizer with its default parameters. $\textbf{H}_0$ and $\textbf{W}_{\eta}$ in influence encoder $\textbf{IE}(\cdot)$ are initialized around $\textbf{0}$ before pre-training and they are fixed during fine-tuning. $\textbf{W}_l$,$\textbf{b}_l$ in $\textbf{MLP}(\cdot)$ (Eq.~\ref{eq:encoder_mlp}) are initialized with identity matrix and $\textbf{0}$, respectively and they are non-trainable until fune-tuning phase. The batch size for pretraining is selected from \{512, 1024\} and the embedding size is 128 for all models. Besides, we use 3 layers for all GNN models, 3 layers for Eq.~\ref{eq:encoder1},\ref{eq:encoder2}, and 2 layers for $\textbf{MLP}(\cdot)$. $\rho \in \{5,15,20\}$ and $\lambda_u$ is chosen from \{1,0.5,0.1\} while $\lambda_p$ is tuned from \{1,0.1,0.01,0.005,0.001\}. And $\lambda_c$ is selected from $\{1e^{-2}, 1e^{-3}, 1e^{-4}\}$. All the temperatures $\tau$ are chosen from \{0.1, 1, 10\}. We use the released code for baselines with grid search and we conduct all the tests on the same device with an NVIDIA GeForce RTX 3090 GPU.

\subsection{Model Utility Analysis (RQ1)}
We commence the evaluation of the unlearning utility of the proposed \model\ paradigm in comparison to the baseline methods. The results are tabulated in Table~\ref{tab:main_tab}, and the following observations can be made based on the Recall@20 and NDCG@20 results.

\begin{itemize}[leftmargin=*]
    \item \textbf{Superior performance retention ability}: After unlearning the well-trained model, our method demonstrates excellent performance retention across multiple datasets compared to retraining. It is noteworthy that on dense datasets such as Movielens, all baselines can maintain relatively good predictive performance. The performance retention capability of the IF-based unlearning method is superior to partition-based methods. This is because partitioning the graph inevitably disrupts its structure, which can have a negative impact on predictive performance.\\\vspace{-0.12in}

    \item \textbf{Drawbacks of partition-based methods}: Traditional partition-based methods partition the graph and, upon receiving an unlearning request, only retrain the affected shards. This approach, which disrupts the graph structure, undoubtedly impacts the predictive performance of GNN and requires time for unlearning. However, within the partition-based category, RecEraser slightly outperforms GraphEraser in terms of performance retention. This is because RecEraser partitions graphs based on edges, which is more reasonable for recommendation systems that prioritize interactive edges. Compared to IF-based methods, partition-based methods may exhibit lower predictive performance, but offer more controllable unlearning effects. These methods require careful training of the affected shards.\\\vspace{-0.12in}
    \item \textbf{Limits for IF-based methods}: IF-based methods calculate unlearning impact through mathematical computations, exhibiting excellent performance retention as unlearning effects can be minimal with little embedding distribution change. Their advantage is computational speed through direct IF calculations. However, this end-to-end estimation may not be accurate for SSL-based GNNs with stochastic structures, leading to influence underestimation and average unlearning effects in practice.
\end{itemize}

\subsection{Model Efficacy Analysis (RQ2)}
We proceed to evaluate and analyze the effects of unlearning, with the experimental results presented in Table~\ref{tab:main_tab}. The following discussions primarily revolve around the MI-BF and MI-NG metrics.

\begin{itemize}[leftmargin=*]
    \item \textbf{Outstanding unlearning efficacy.}: 
    Based on MI-BF and MI-NG metrics, \model\ shows strong unlearning results. Across datasets and backbones, \model\ achieves MI-BF and MI-NG values above 1. MI-BF > 1 shows reduced probability of recommending edges marked for unlearning versus before. Similarly, MI-NG > 1 means unlearned edges are less likely to be recommended than negative samples, protecting user privacy after unlearning. This shows \model\ effectively simulates edge removal through pre-training and identifies these effects via fine-tuning. \model\ works with various SSL-based GNNs and matches each model's embedding distribution through unlearning (detailed later). This comes from using $\mathcal{L}_M$ to capture loss functions, training methods, random processes, and network structure of each GNN, yielding similar embedding distributions.
    \\\vspace{-0.12in}

    \item \textbf{Limitations of partition-based methods.}: As mentioned earlier, partitioning the graph disrupts its overall integrity, leading to negative impacts on predictive performance after unlearning. Additionally, many SSL-based methods are designed to address the sparsity issue in recommendation systems, where the interaction graph is sparse. Consequently, in SSL-based GNNs, when an unlearning request is made, almost all edges and nodes are more or less affected since they are involved in the formation and maintenance of self-supervised labels. Therefore, training only a subset of shards is unlikely to yield satisfactory unlearning results.Specifically, it is challenging to achieve MI-NG > 1.5.
    \\\vspace{-0.12in}
    \item \textbf{Ineffectiveness of IFs for SSL-based GNNs.}: 
    IF-based methods need complex mathematical designs. But training SSL-based GNNs is more random and unpredictable. SSL-based GNNs use random network structures and signals, making gradient estimation hard. This makes IF methods less effective for SSL-based GNNs, leading to poor unlearning. Table 2 shows that with SimGCL and SGL backbones, CEU and GIF perform poorly on MI-NG, sometimes below 1. This means that while unlearning reduces recommendations of learned edges, these rates stay higher than for negative samples. This suggests learned edges were once positive samples, risking user data leaks.
\end{itemize}

\subsection{Ablation Study (RQ3)}
\begin{table}[t]
    \centering
    \caption{Ablation study of \model\ based on SimGCL.}
    \label{tab:ablation_study}
    \small
    \setlength{\tabcolsep}{0.6mm}
    \vspace{-0.1in}

\begin{tabular}{|c|cccc|cccc|}
\hline
Datasets          & \multicolumn{4}{c|}{Movielens-1M}                                                                & \multicolumn{4}{c|}{Gowalla}                                                                      \\ \hline
Metrics           & \multicolumn{1}{c|}{Recall} & \multicolumn{1}{c|}{NDCG}   & \multicolumn{1}{c|}{MI-BF}  & MI-NG  & \multicolumn{1}{c|}{Recall}  & \multicolumn{1}{c|}{NDCG}   & \multicolumn{1}{c|}{MI-BF}  & MI-NG  \\ \hline
- $\mathcal{L}_u$ & \multicolumn{1}{c|}{0.2338} & \multicolumn{1}{c|}{0.3331} & \multicolumn{1}{c|}{1.5347} & 0.5948 & \multicolumn{1}{c|}{0.2543}  & \multicolumn{1}{c|}{0.1677} & \multicolumn{1}{c|}{1.3452} & 0.5639 \\ \hline
- $\mathcal{L}_p$ & \multicolumn{1}{c|}{0.1863} & \multicolumn{1}{c|}{0.2682} & \multicolumn{1}{c|}{4.8245} & 3.7239 & \multicolumn{1}{c|}{0.17107} & \multicolumn{1}{c|}{0.1117} & \multicolumn{1}{c|}{6.8021} & 3.3367 \\ \hline
- $\mathcal{L}_c$ & \multicolumn{1}{c|}{0.2300} & \multicolumn{1}{c|}{0.3262} & \multicolumn{1}{c|}{3.0013} & 2.1789 & \multicolumn{1}{c|}{0.2507}  & \multicolumn{1}{c|}{0.1633} & \multicolumn{1}{c|}{4.1979} & 1.6625 \\ \hline
- FineTune        & \multicolumn{1}{c|}{0.2211} & \multicolumn{1}{c|}{0.3188} & \multicolumn{1}{c|}{2.2226} & 1.6998 & \multicolumn{1}{c|}{0.2398}  & \multicolumn{1}{c|}{0.1563} & \multicolumn{1}{c|}{3.3232
} & 1.3047 \\ \hline
Ours              & \multicolumn{1}{c|}{0.2336} & \multicolumn{1}{c|}{0.3323} & \multicolumn{1}{c|}{3.1981} & 2.3575 & \multicolumn{1}{c|}{0.2540}  & \multicolumn{1}{c|}{0.1670} & \multicolumn{1}{c|}{4.3520} & 1.8075 \\ \hline
\end{tabular}

    \vspace{-0.15in}
\end{table}

We conducted the ablation experiments on the Movielens and Gowalla datasets using the SimGCL model as the backbone and made the following findings. The results are shown in Table~\ref{tab:ablation_study}
\begin{itemize}[leftmargin=*]
    \item $\textbf{-}\mathcal{L}_u$:
    When we remove the loss function $\mathcal{L}_u$, we observe that the model's predictive performance remains relatively unchanged in terms of Recall and NDCG. However, there is a significant drop in the unlearning efficacy, specifically MI-BF and MI-NG. This is because $\mathcal{L}_u$ plays a role in reducing the probability of recommending edges that need to be unlearned. Nevertheless, we found that even without $\mathcal{L}_u$, MI-BF still exceeds 1, indicating a decrease in the probability of recommending the unlearned edges. This is due to the presence of the BPR loss in $\mathcal{L}_M$, which contributes to reducing the similarity between the two nodes in the interaction pairs that need to be unlearned. However, solely relying on this BPR loss is insufficient to eliminate the effects caused by the unlearning edges. Therefore, MI-NG remains below 1, indicating that the probability of recommending unlearned edges is still higher than that of negative samples.\\\vspace{-0.12in}
    \item $\textbf{-}\mathcal{L}_p$: 
    The purpose of $\mathcal{L}_p$ is to preserve useful information from the original trained model to some extent. Its goal is to prevent the model from deviating too far from the original embedding distribution during the pre-training process, where the edges requiring unlearning are randomly simulated. After all, in practice, the edges that need to be unlearned typically constitute a small portion of all edges. Removing $\mathcal{L}_p$ significantly compromises the predictive performance during the unlearning process, even though the unlearning effects, such as MI-BF and MI-NG, may increase. However, this trade-off is not worth it.\\\vspace{-0.12in}
    \item $\textbf{-}\mathcal{L}_c$: The influence of $\mathcal{L}_c$ on the overall outcome, whether it is utility or unlearning efficacy, is relatively small. $\mathcal{L}_c$ can be considered as a regularization term, primarily used to smooth the randomly simulated unlearning requests during the pre-training process. Therefore, the weight of $\mathcal{L}_c$, denoted as $\lambda_c$, does not need to be very large. To some extent, the smoothing effect of $\mathcal{L}_c$ on unlearning requests may slightly decrease the probability of recommending the interactions similar to unlearning requests. However, generally speaking, these interactions are still regarded as positive pairs by the recommendation system and are still likely to be recommended like other normal positive instances.\\\vspace{-0.12in}

    \item $\textbf{-}FineTune$: In our experiment of adversarial attacks, we found that even without fine-tuning the pre-trained influence encoder (IE), we could still achieve good results (both in utility and efficacy) when unlearning requests were received and directly applied to unlearn using \textbf{IE}. This indicates that our pre-trained \textbf{IE} is capable of effectively identifying and mitigating the impact of adversarial edges on embeddings. In fact, we can consider the $\Delta\bar{\textbf{E}}_0$ in \textbf{IE} as learnable noises. When unlearning requests arrive, the embeddings of affected nodes are first masked by the noises $\Delta\bar{\textbf{E}}_0$ in \textbf{IE}, and then fine-tuned to adapt to the specific contextual changes brought by the unlearning requests. To put it simply, the pre-training stage is analogous to the noise injection process of a diffusion model, while the fine-tuning stage is similar to the reconstruction and generation process of diffusion~\cite{croitoru2023diffusion,ho2020denoising,sohl2015deep}.
    
\end{itemize}

\subsection{Embedding Distribution Shift Study (RQ4)}
\begin{figure*}[t]
    \centering
    \includegraphics[width=0.45\textwidth]{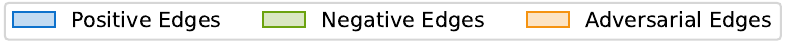}\vspace{-0.05in}\\
    \subfigure[Prediction distribution of GCN after unlearning on Movielens-1M dataset]{
        \hspace{-0.003\textwidth}\includegraphics[width=0.125\textwidth]{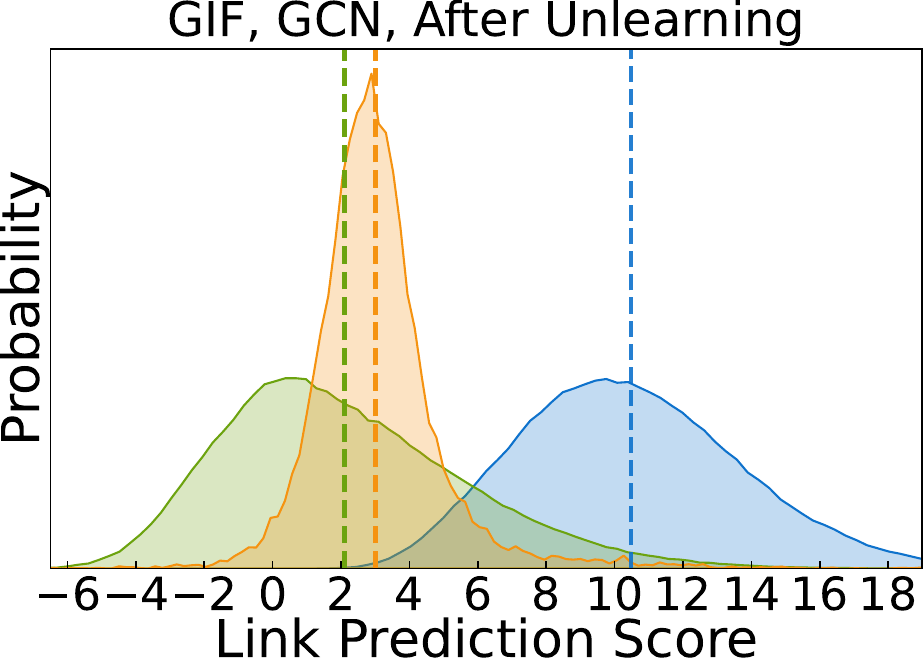}
        \hspace{-0.005\textwidth}\includegraphics[width=0.125\textwidth]{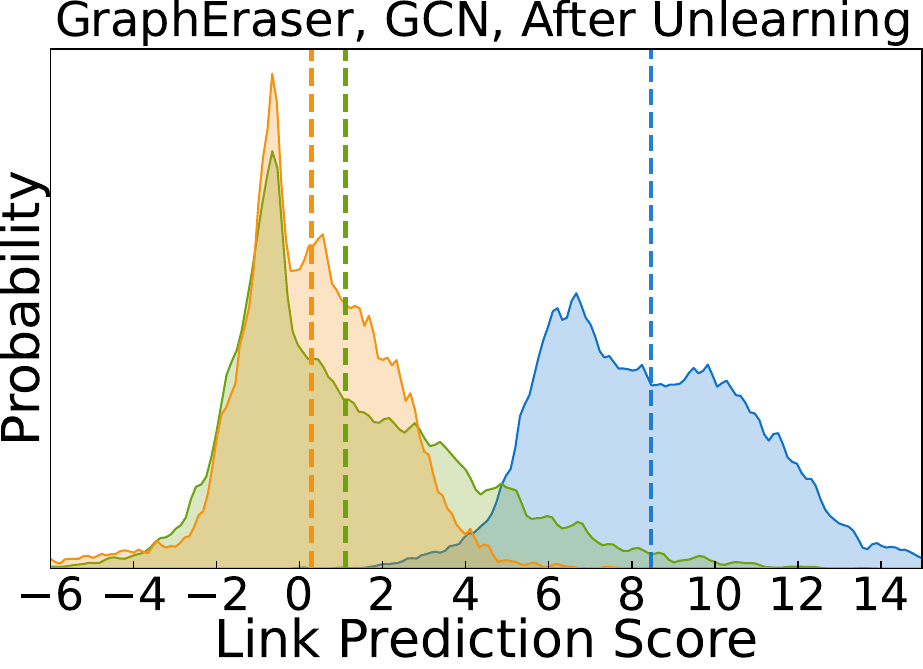}
        \hspace{-0.005\textwidth}\includegraphics[width=0.125\textwidth]{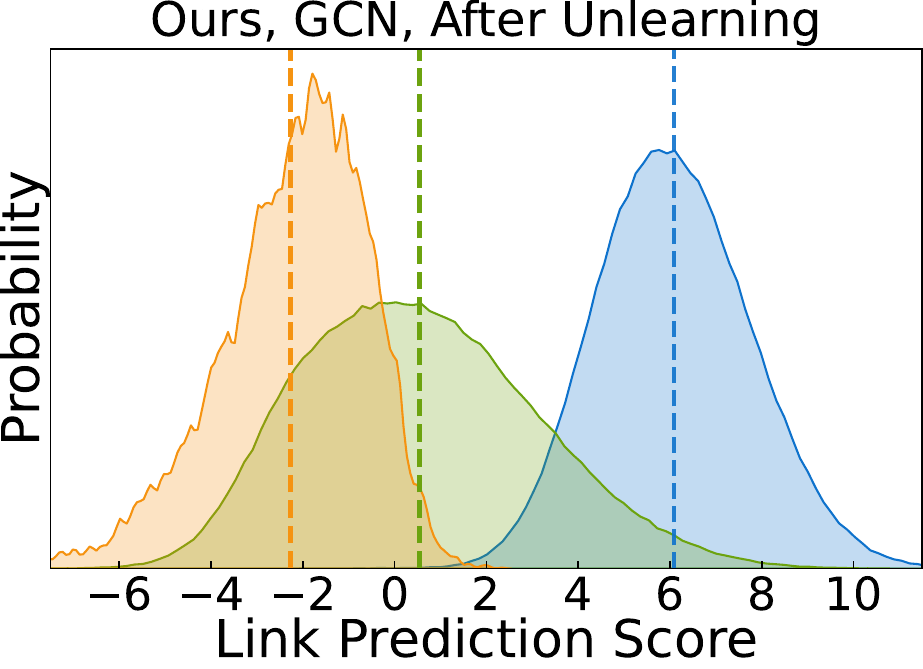}
        \hspace{-0.005\textwidth}\includegraphics[width=0.125\textwidth]{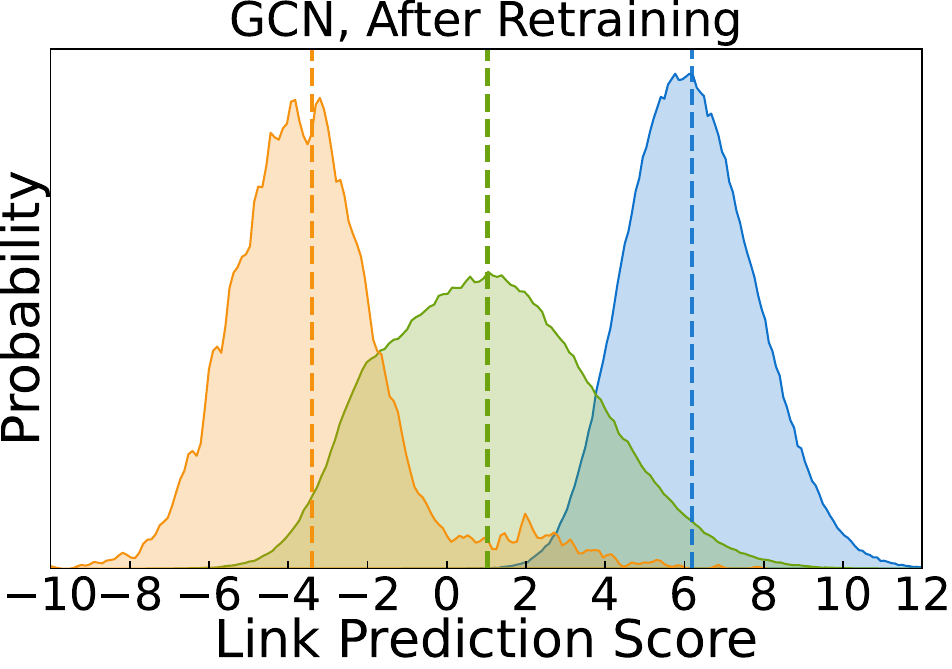}
    }%
    \subfigure[Prediction distribution of SGL after unlearning on Movielens-1M dataset]{
        \hspace{-0.005\textwidth}\includegraphics[width=0.125\textwidth]{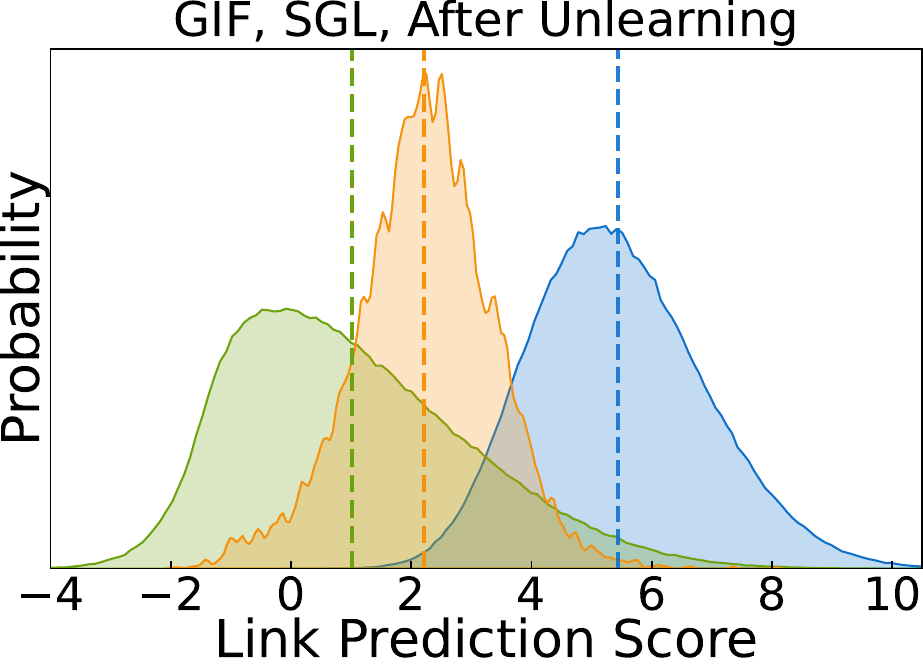}
        \hspace{-0.005\textwidth}\includegraphics[width=0.125\textwidth]{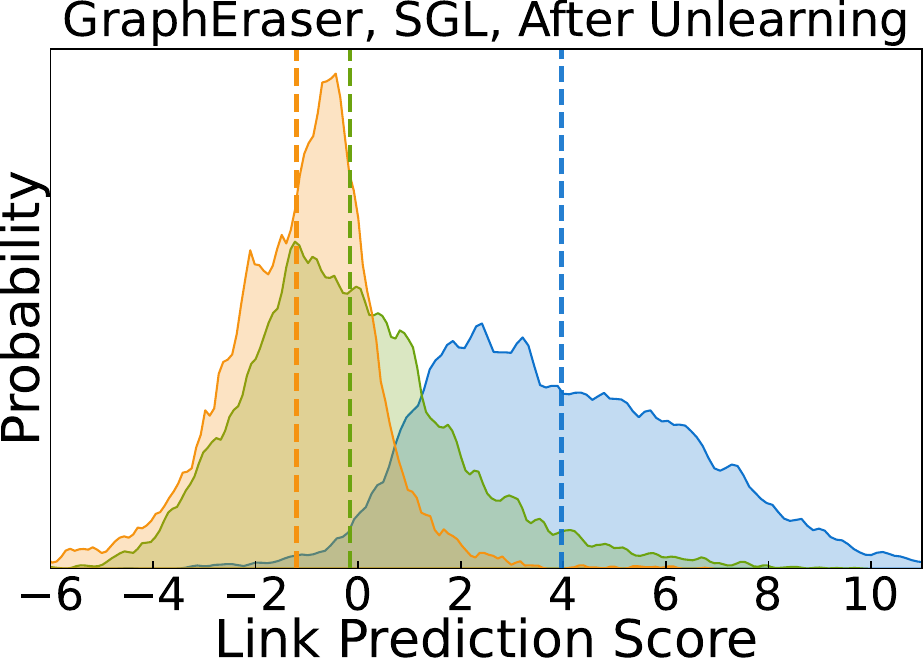}
        \hspace{-0.005\textwidth}\includegraphics[width=0.125\textwidth]{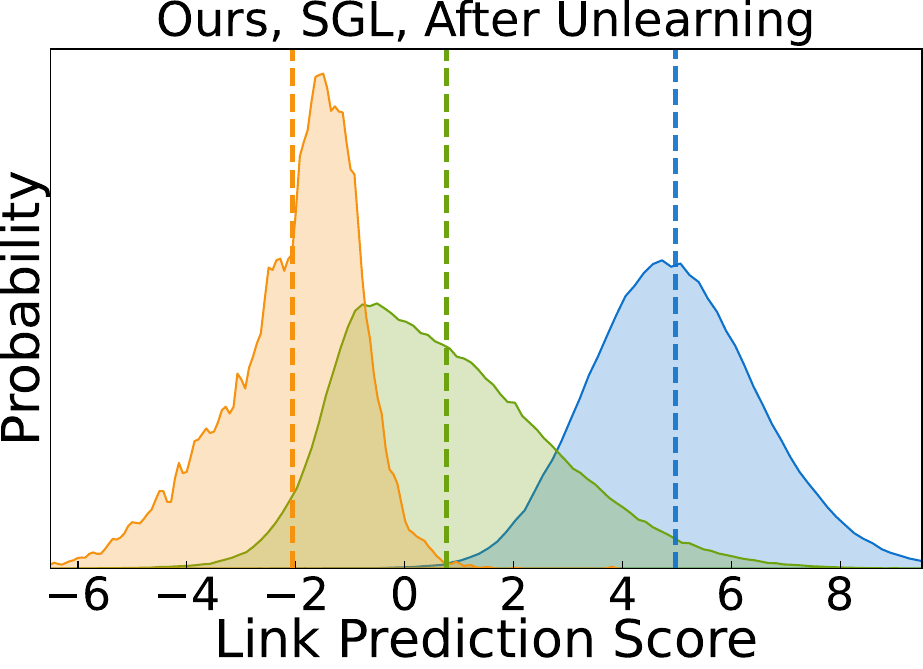}
        \hspace{-0.005\textwidth}\includegraphics[width=0.125\textwidth]{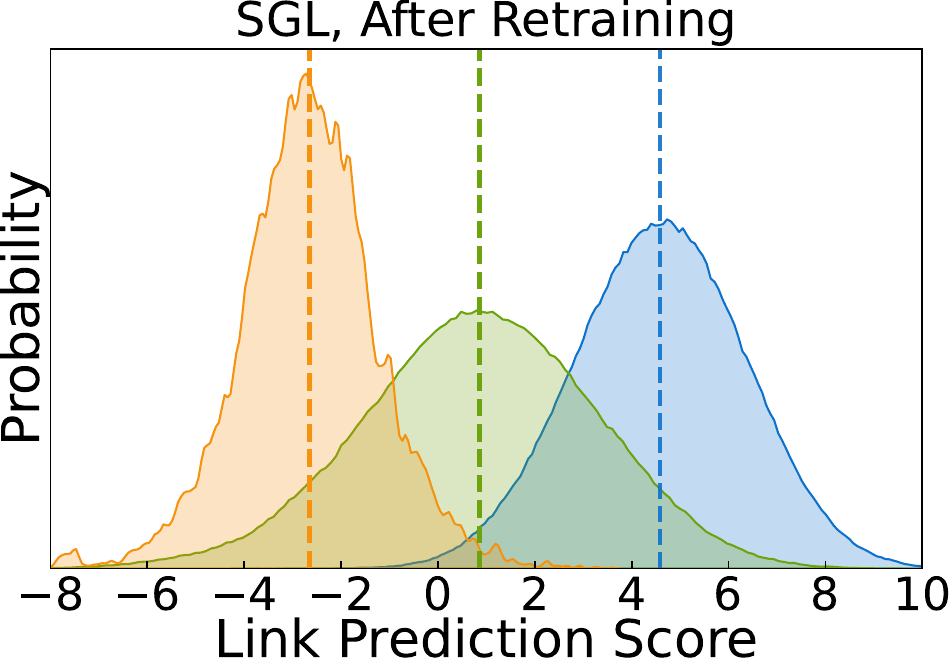}
    }\vspace{-0.1in}
    
    \subfigure[Prediction distribution of SGL after unlearning on Yelp2018 dataset]{
        \hspace{-0.003\textwidth}\includegraphics[width=0.125\textwidth]{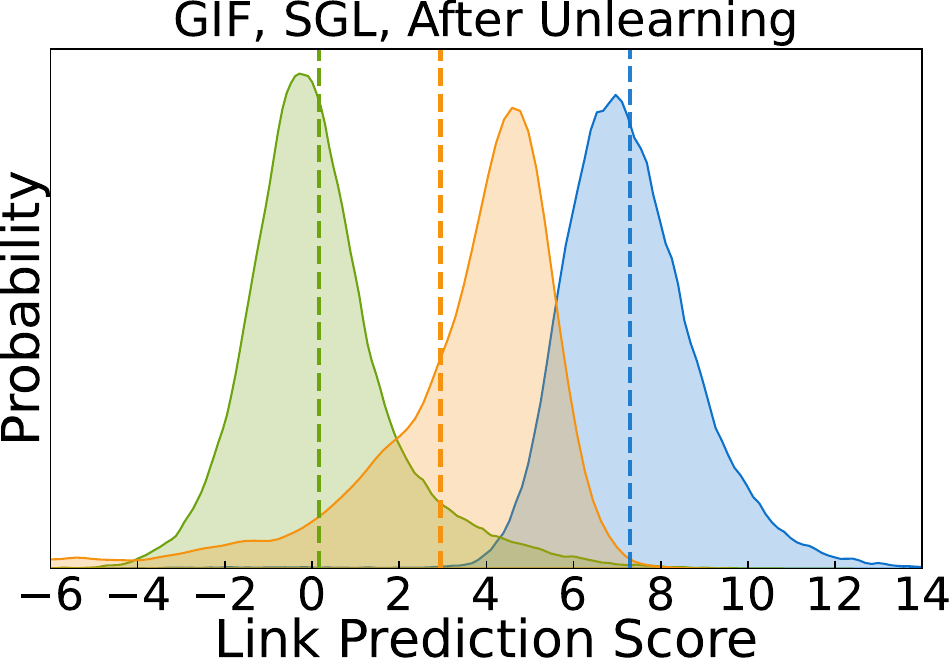}
        \hspace{-0.005\textwidth}\includegraphics[width=0.125\textwidth]{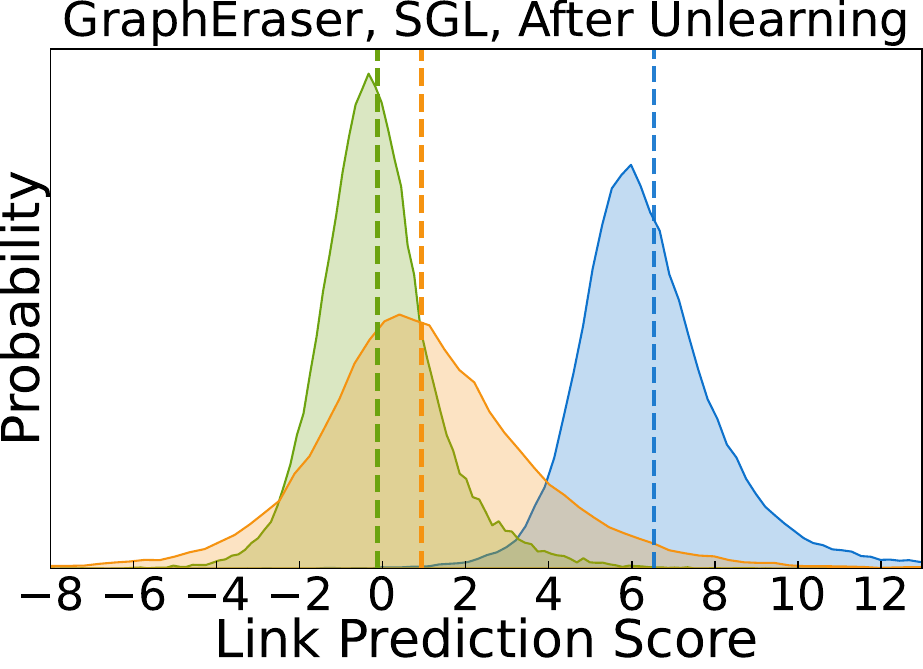}
        \hspace{-0.005\textwidth}\includegraphics[width=0.125\textwidth]{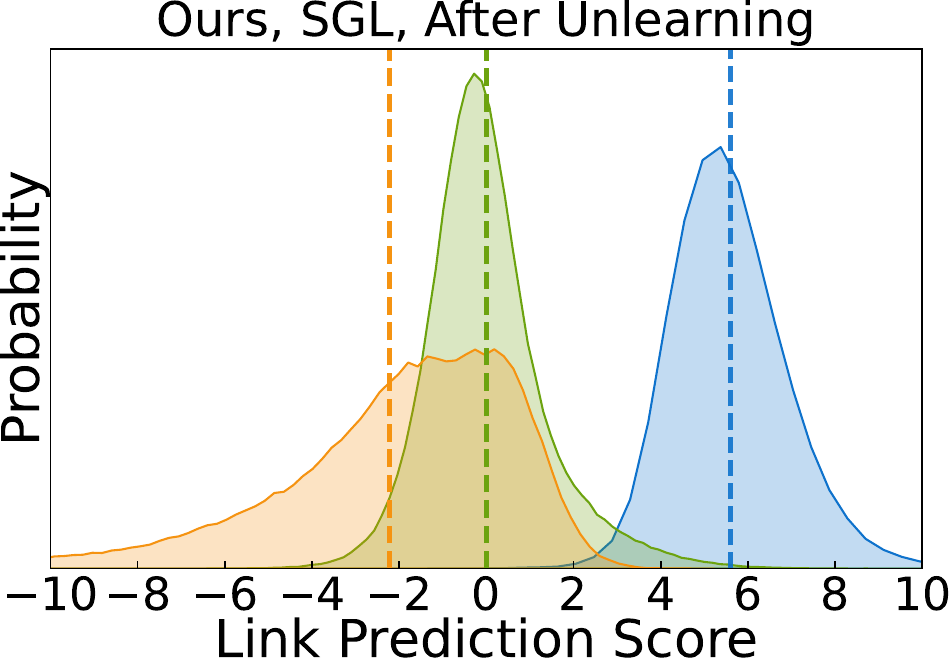}
        \hspace{-0.005\textwidth}\includegraphics[width=0.125\textwidth]{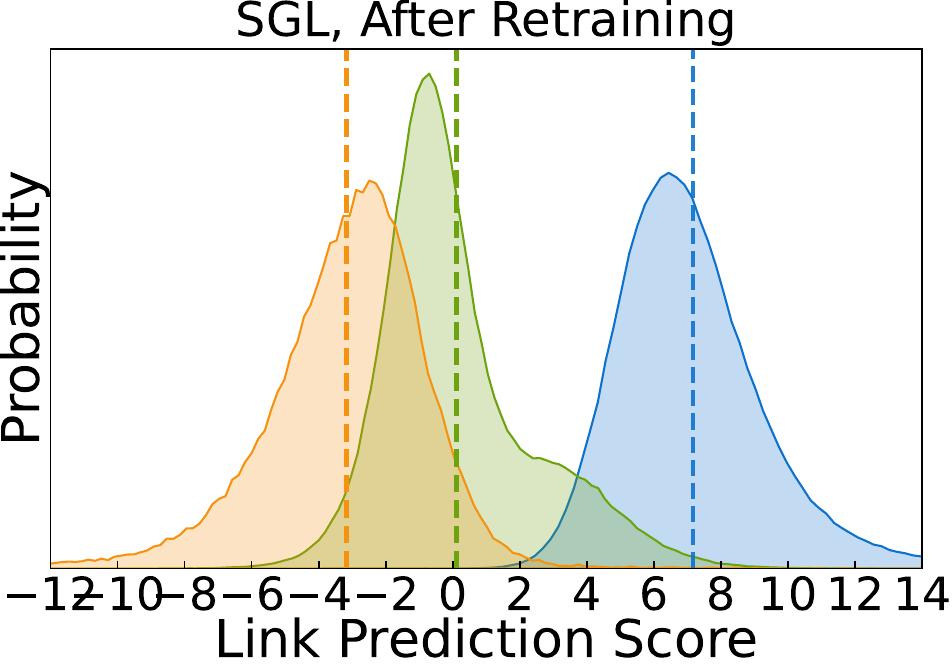}
    }%
    \subfigure[Prediction distribution of SimGCL after unlearning on Yelp2018 dataset]{
        \hspace{-0.005\textwidth}\includegraphics[width=0.125\textwidth]{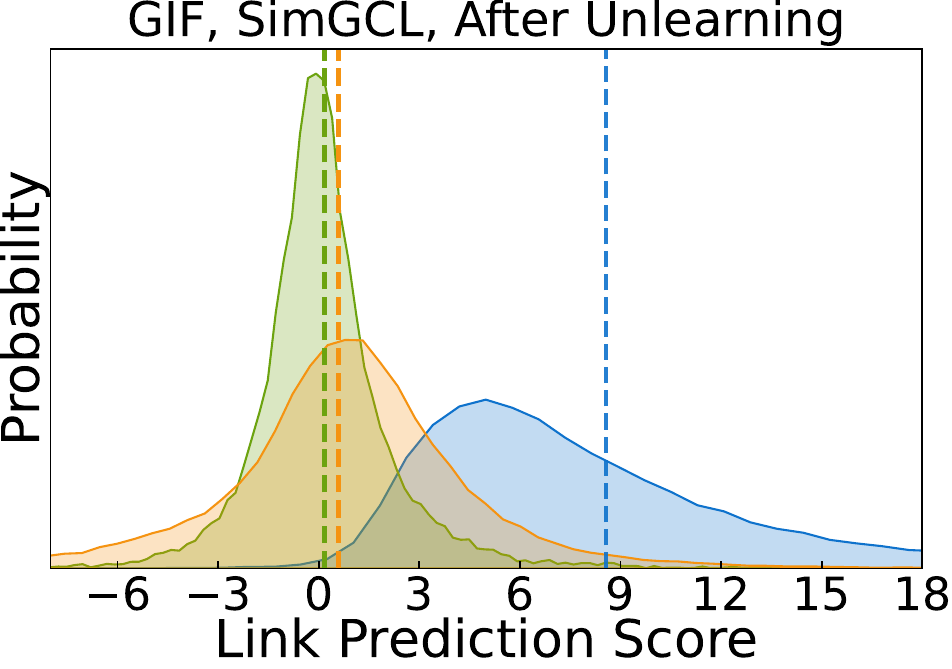}
        \hspace{-0.005\textwidth}\includegraphics[width=0.125\textwidth]{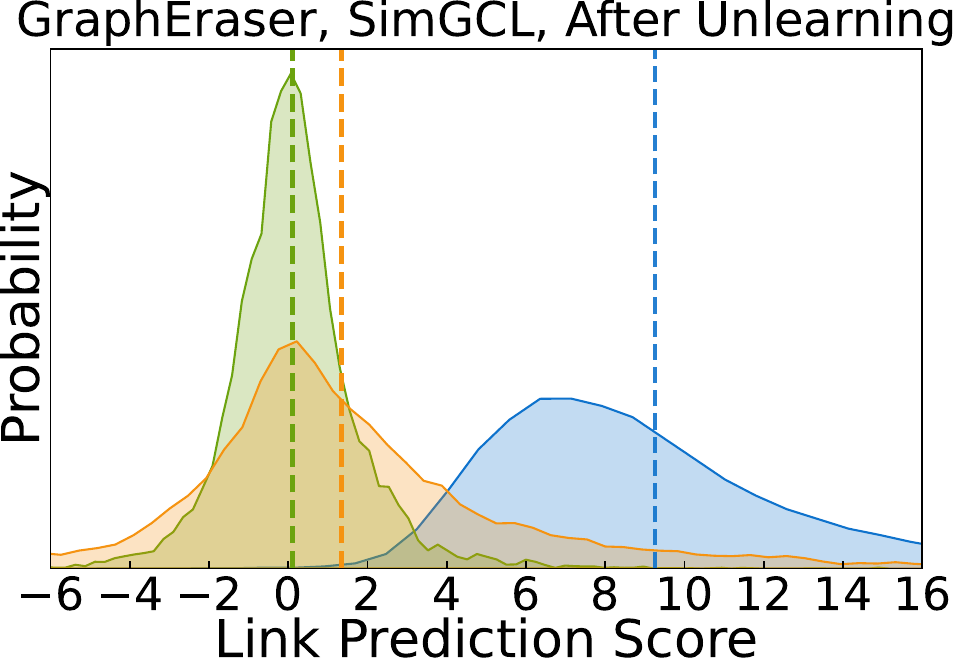}
        \hspace{-0.005\textwidth}\includegraphics[width=0.125\textwidth]{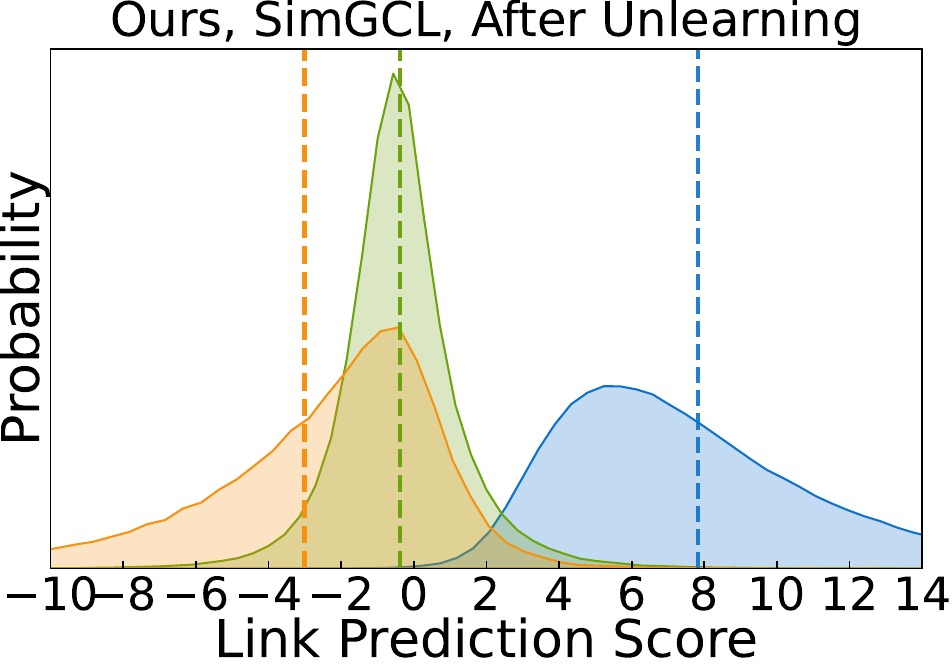}
        \hspace{-0.005\textwidth}\includegraphics[width=0.125\textwidth]{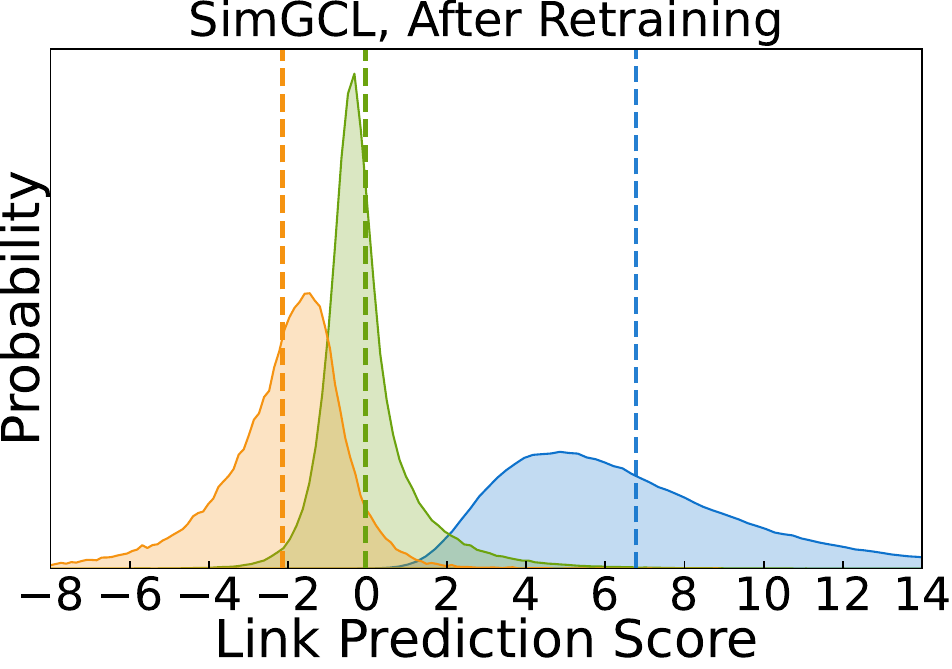}
    }
    \vspace{-0.15in}
    \caption{Visualization for the predictions of positive, negative, and adversarial edges in the Movielens-1M and Yelp2018 datasets. Four methods with three backbone models are compared, including the IF-based method GIF, the partition-based method GraphEraser, the retraining-based exact unlearning, and our proposed \model, on the GCN, SGL, SimGCL backbones.}
    \label{fig:unlearning_efficacy}
    \vspace{-0.1in}
\end{figure*}
\begin{figure}[t]
    \centering
    \subfigure[Distribution on ML-1M before unleaning]{
        \hspace{-0.003\textwidth}\includegraphics[width=0.125\textwidth]{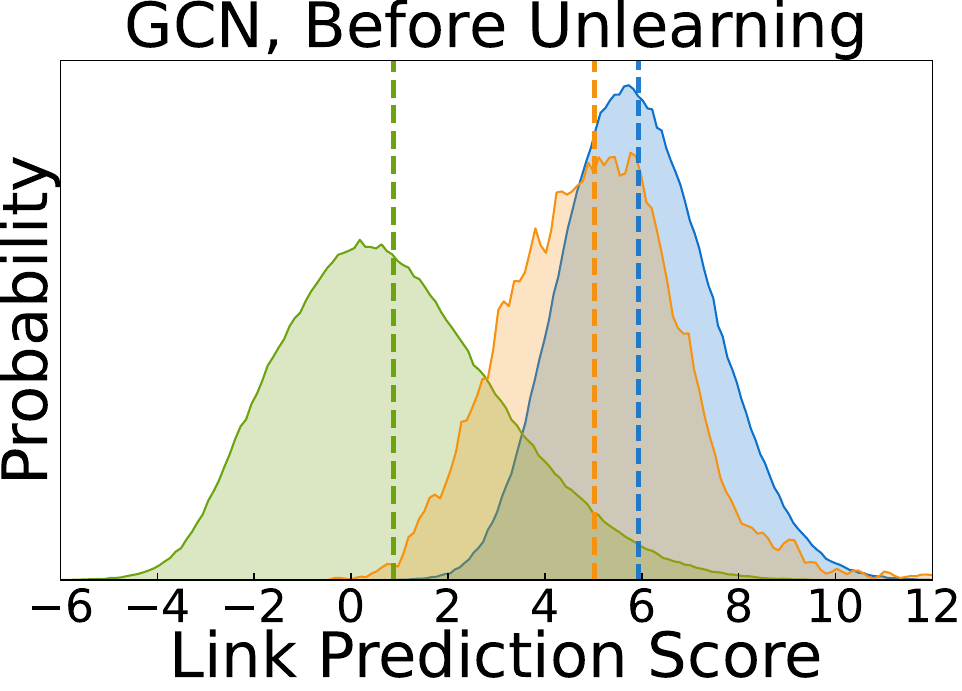}
        \hspace{-0.005\textwidth}\includegraphics[width=0.125\textwidth]{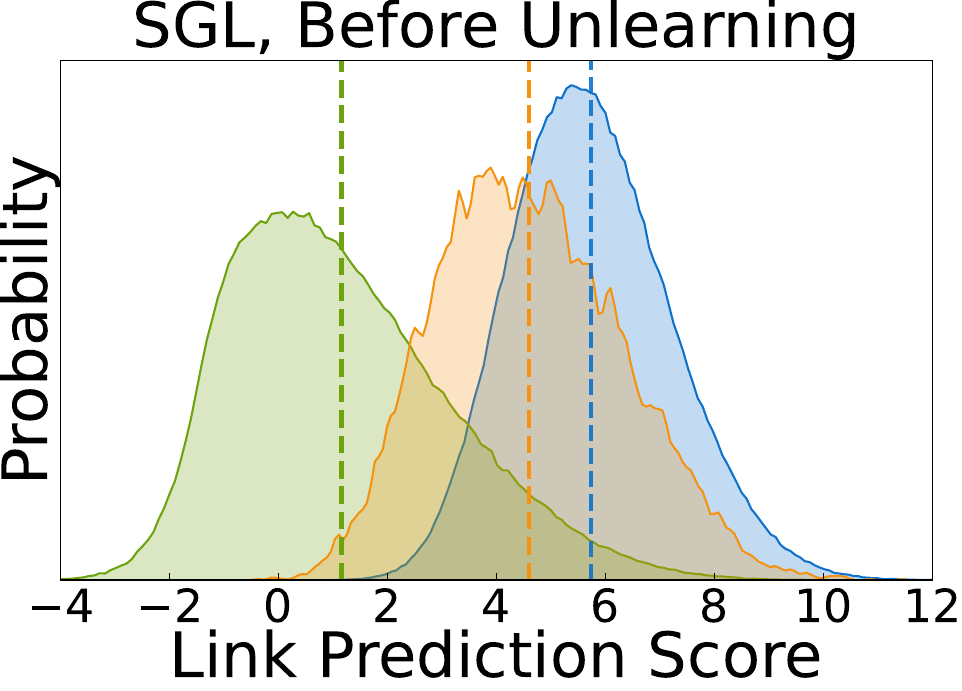}
    }%
    \subfigure[Distribution on Yelp before unleaning]{
        \hspace{-0.005\textwidth}\includegraphics[width=0.125\textwidth]{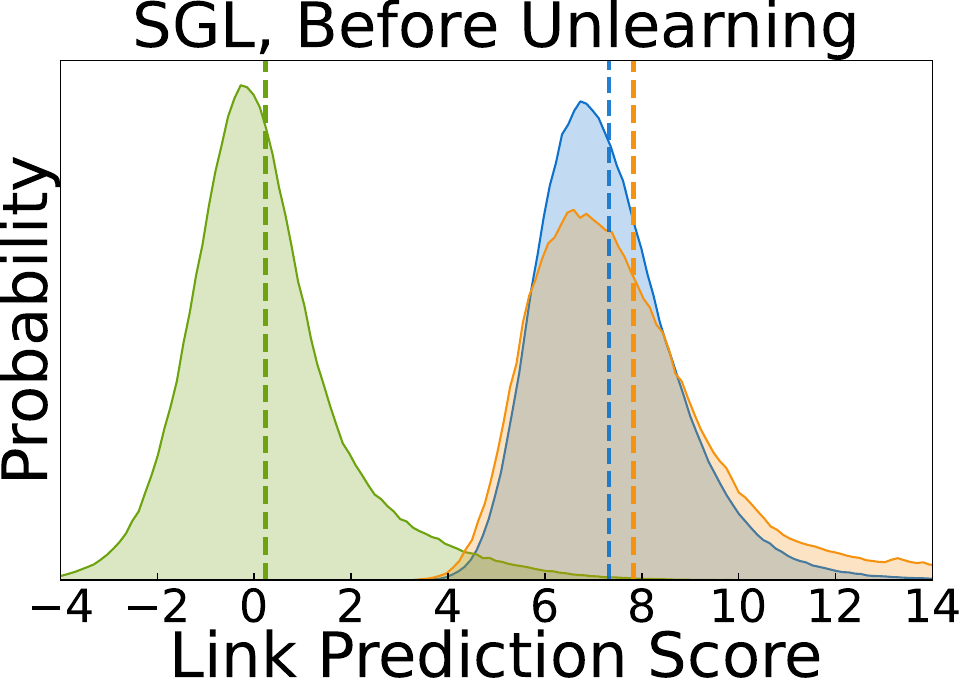}
        \hspace{-0.005\textwidth}\includegraphics[width=0.125\textwidth]{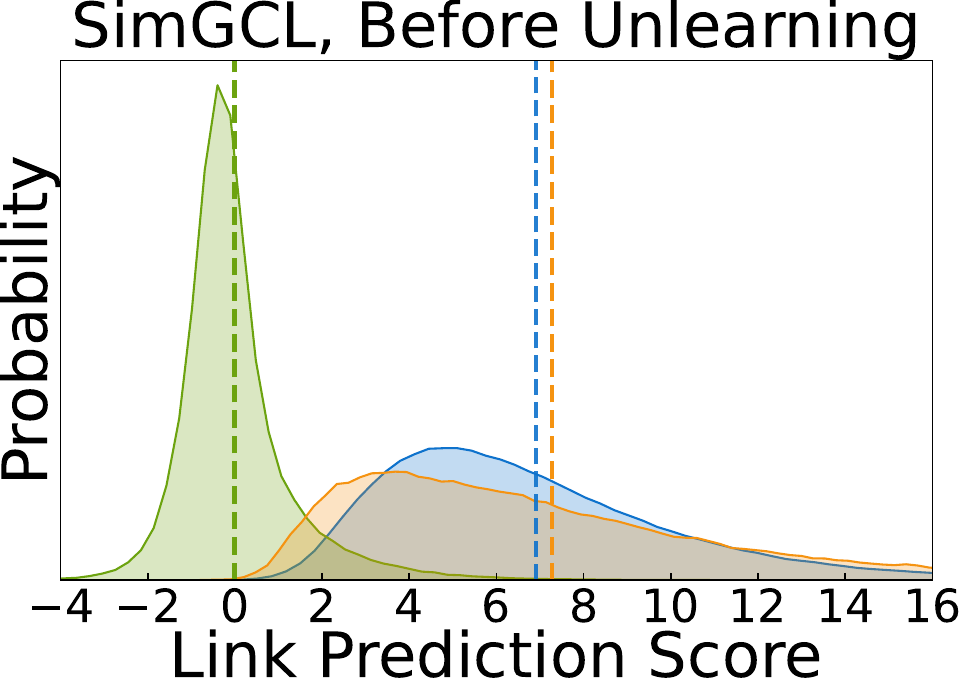}
    }
    
    \vspace{-0.2in}
    \caption{Visualization of the prediction distributions for Movielens-1M and Yelp2018 datasets before unlearning.}
    \label{fig:before_unlearning}
    \vspace{-0.12in}
\end{figure}

In this section, we mainly discuss the changes in the embedding distribution of various recommendation system models before and after unlearning, as well as their comparison with the results obtained from retraining. The predicted score distribution before unlearning is shown in Fig.~\ref{fig:before_unlearning}, while the distribution after unlearning is shown in Fig.~\ref{fig:unlearning_efficacy}. The dashed lines in the figures represent the mean predicted scores for the corresponding types of edges. Based on these results, we have the following observations.
\begin{itemize}[leftmargin=*]
    \item \textbf{Adversarial Edges}. Before unlearning, the predicted scores for all adversarial edges in various models are close to positive instances, which is reasonable since they were labeled as positive during the training process. These Adversarial Edges are sampled based on a set of items that a trained GCN predicts as the least likely to be interacted by users. Therefore, after removing the adversarial edges and retraining, we found that the mean predicted scores for them are lower than regular negative instances.\\\vspace{-0.12in}
    \item \textbf{P-based Methods vs. IF-based Methods}. 
    In general, P-based methods demonstrate better controllability and efficacy in terms of unlearning efficacy compared to IF-based methods. From Fig.~\ref{fig:unlearning_efficacy}, it can be observed that the mean predicted scores for adversarial edges given by GraphEraser are more closer to the mean of negative instances, while in many backbones, GIF still exhibits significantly higher predicted scores for adversarial edges compared to negative edges (e.g., GIF for SGL in both datasets). Although GraphEraser outperforms GIF in terms of unlearning efficacy, it still falls short of meeting the requirements, as evident in the picture of the SGL on the Yelp dataset, where the scores for adversarial edges still remain higher than negative instances. This implies that even after unlearning, it is still possible to infer the existence of these edges based on the embeddings.\\\vspace{-0.12in}
    \item \textbf{Our \model~Paradigm}. Our method demonstrates good unlearning efficacy on both datasets and across three backbones. In all experiments, after unlearning, the mean predicted scores for adversarial edges are significantly lower than negative instances. Furthermore, the score distribution of adversarial edges is enveloped by the score distribution of negative instances, ensuring minimal information leakage about these edges after unlearning. In other words, it becomes impossible to infer the existence of these edges based on the embeddings. Visually, the embedding distribution of our model closely resembles that of retraining.\\\vspace{-0.12in}
\end{itemize}

\begin{figure}[t]
    \centering
    \subfigure[GPU memory usage on Movielens-1M ]{
        \includegraphics[width=0.23\textwidth]{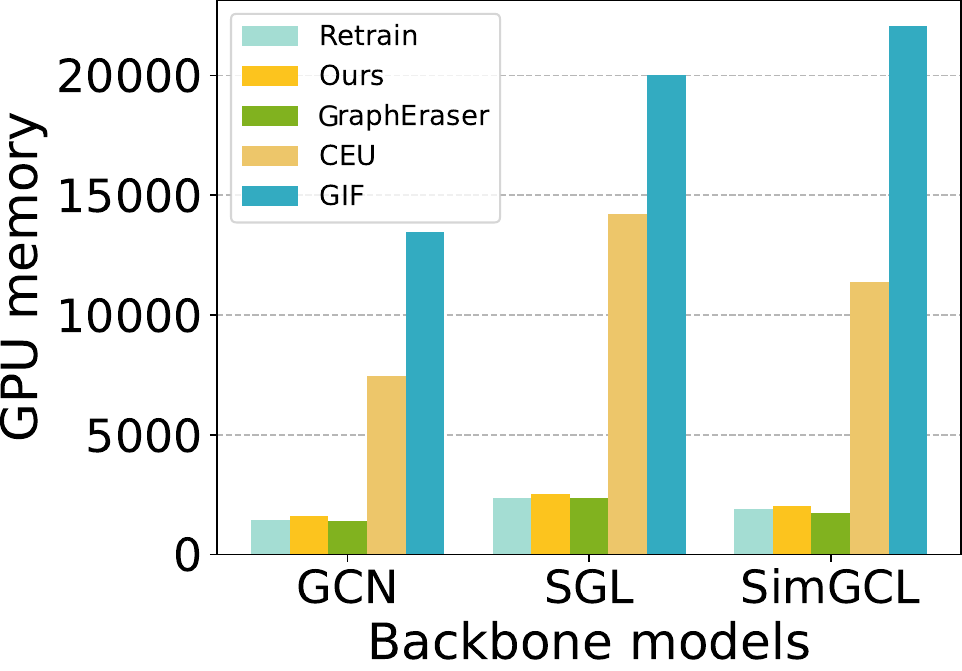}
    }\subfigure[GPU memory usage on Gowalla ]{
        \includegraphics[width=0.23\textwidth]{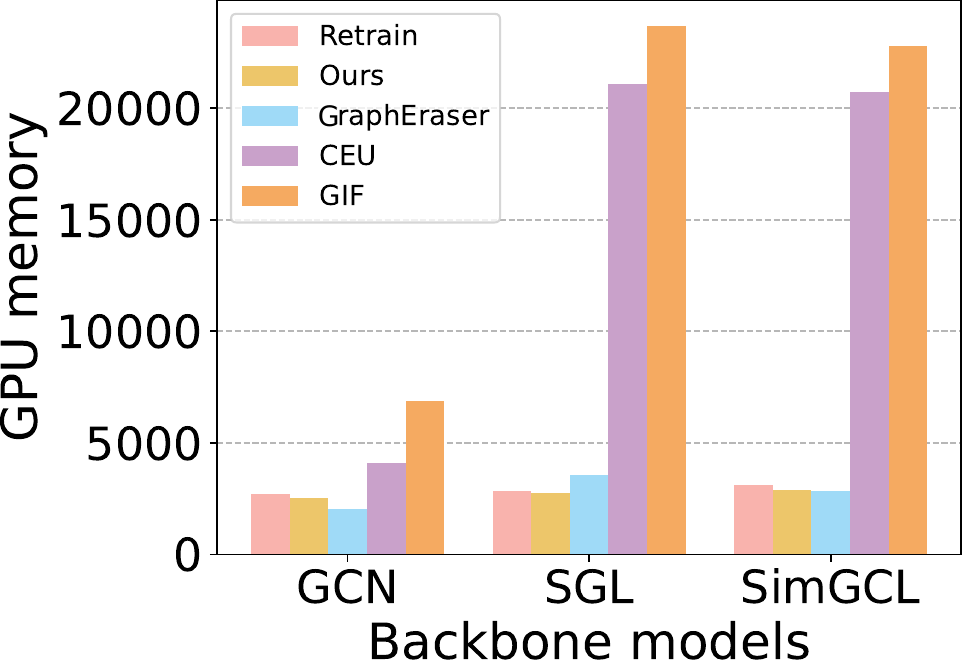}
    }
    \vspace{-0.2in}
    \caption{Comparison of GPU memory usage with different methods on various backbones based on two datasets.}
    \label{fig:mem}
    \vspace{-0.1in}
\end{figure}

\subsection{Unlearning Efficiency Analysis (RQ5)}
\begin{table}[t]
\setlength{\tabcolsep}{0.5mm}

\caption{The time for processing unlearning requests of different methods based on the SimGCL model and three datasets.}
\label{tab:time}
\vspace{-0.1in}
\resizebox{0.49\textwidth}{!}{
\begin{tabular}{c|c|c|c|c|c|c|c|c}
   \toprule[1.1pt]
   Dataset	&Retrain	&GraphEraser &GIF &CEU &Ours(0)	&Ours(1) &Ours(2) &Ours(3) \\
   \midrule
   ML-1M &1208.354	&993.064	&8.278	&6.943	&0.312	&7.034	&14.221	&21.517 \\
   Yelp2018  &3029.606	&1834.413	&146.384	&52.216	&0.468	&19.71	&39.201	&56.023 \\
   Gowalla &839.07	&795.995 &131.753 &39.089	&0.432	&3.37	&6.734	&10.026 \\
   \bottomrule[1.1pt]
\end{tabular}}
\vspace{-0.15in}
\end{table}

\begin{figure}[t]
    \centering
    \subfigure[Unlearning efficacy and performance changes during pre-training]{
        \includegraphics[width=0.21\textwidth]{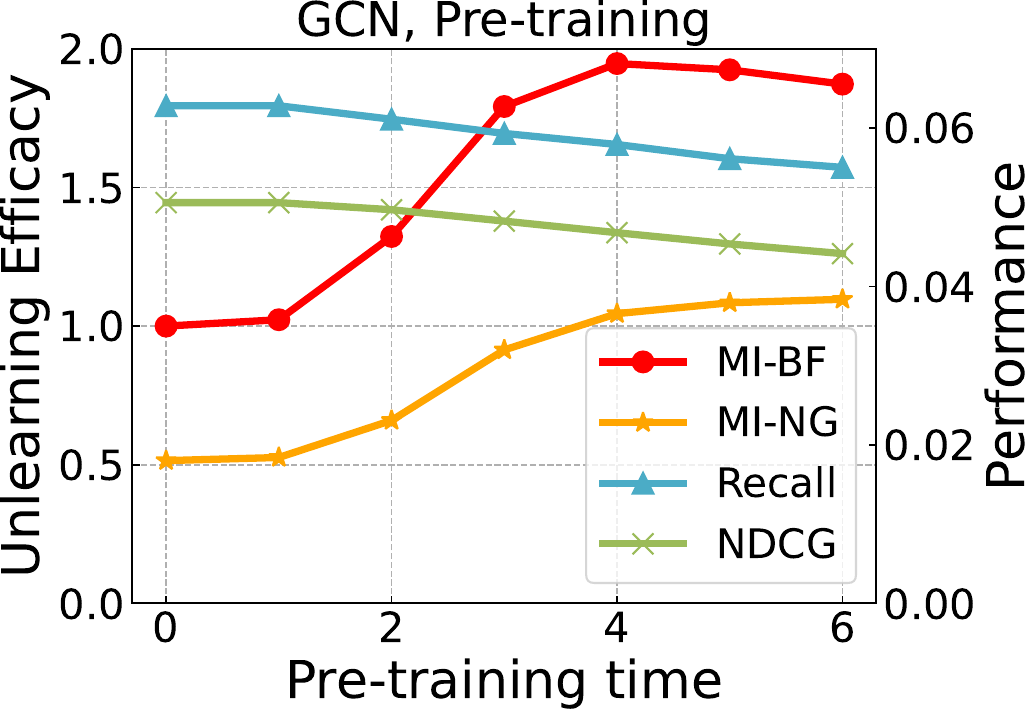}
        \hspace{0.005\textwidth}\includegraphics[width=0.21\textwidth]{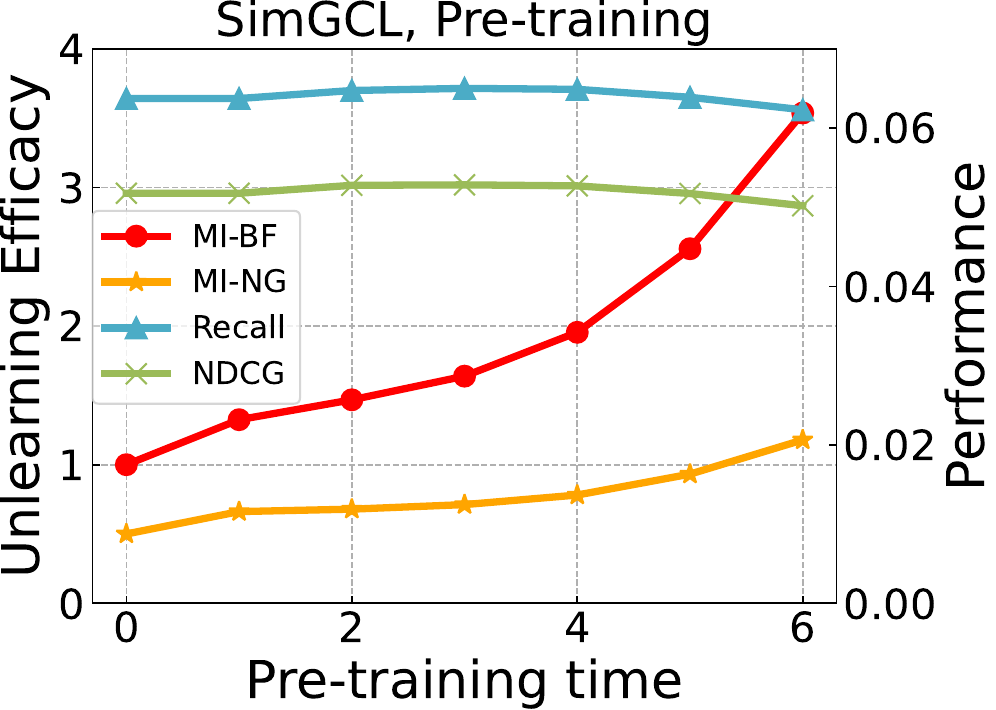}
    }\vspace{-0.1in}
    \subfigure[Unlearning efficacy and performance changes during fine-tuning]{
        \includegraphics[width=0.21\textwidth]{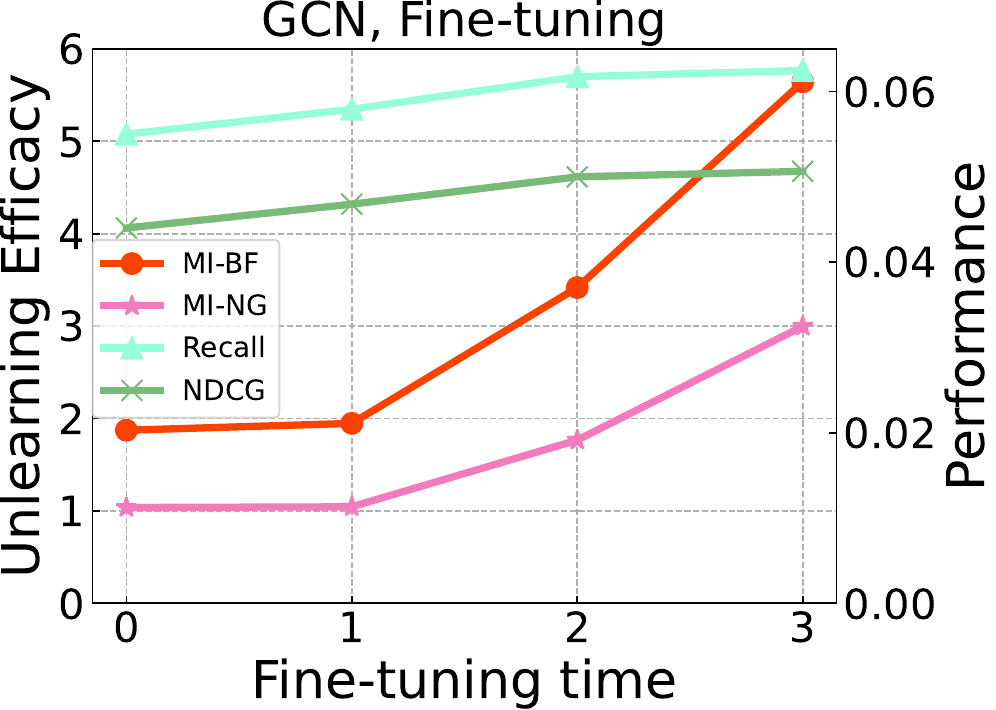}
        \hspace{0.005\textwidth}\includegraphics[width=0.21\textwidth]{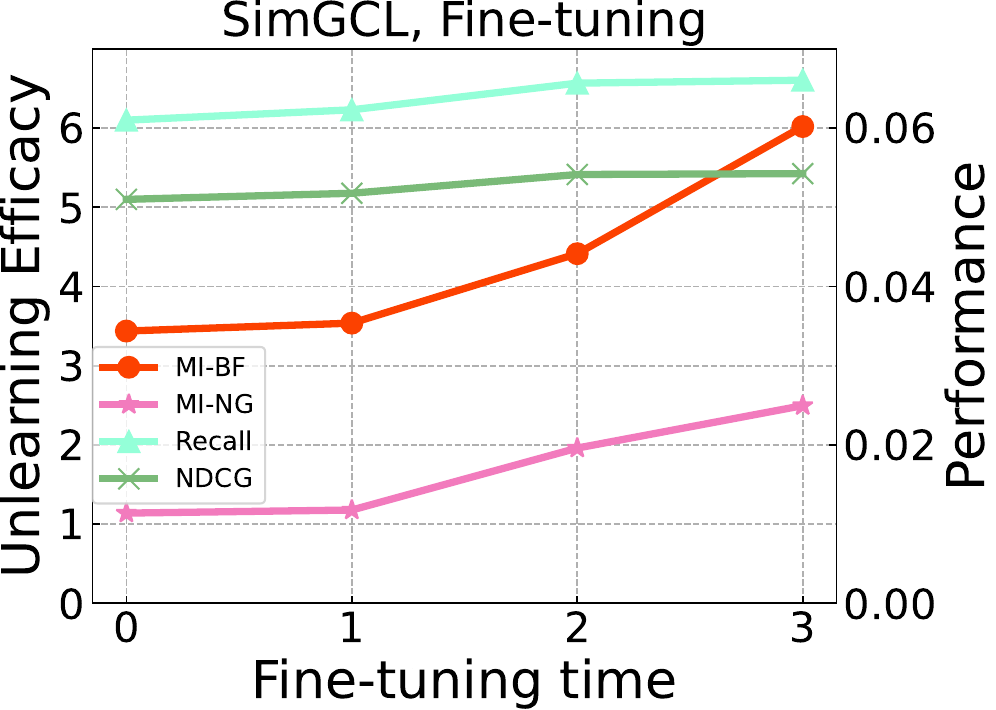}
    }\vspace{-0.18in}    
    \caption{Unlearning efficacy and performance v.s. pre-training and fine-tuning processes of \model.}
    
    \label{fig:pretrain-finetune}
    \vspace{-0.12in}
\end{figure}

To assess the efficiency of our \model\ paradigm, we conduct a comparative analysis of various models in terms of GPU memory and processing time when unlearning requests come. Fig.~\ref{fig:mem} shows and GPU memory cost and Tab.~\ref{tab:time} shows the processing time, in which \textit{Ours(0)} means no finetuning and \textit{Ours(1)} denotes funetuning for 1 epoch. And we have following discussions.
\begin{itemize}[leftmargin=*]
    \item \textbf{High Memory Cost for IF-based Methods}. 
    IF methods typically require second-order gradients (Hessian matrix) to estimate unlearning effects. This requires storing computation graphs and using the first-order graph to calculate Hessian, consuming significant GPU memory. With sufficient GPU memory for the full dataset's second-order graph, unlearning is fast (like GIF/CEU on Movielens). Without sufficient memory, batching becomes necessary, increasing time and reducing IF accuracy. CEU utilizes HVP~\cite{pearlmutter1994fast} and CG~\cite{steihaug1983conjugate} to estimate inverse Hessian, saving some computational resources. However, it still requires storing graphs, which are complex due to SSL-based GNNs' stochastic structures. This increases computational complexity and reduces estimation accuracy of IF-based unlearning methods.
    \\\vspace{-0.12in}
    \item \textbf{High Time Cost for P-based Methods}. P-based methods require retraining the shards affected by unlearning requests and then aggregating all the shards. This process takes time, especially when unlearning requests involve a large number of shards. P-based methods are relatively efficient in terms of GPU memory consumption. Their GPU consumption is comparable to retraining, or sometimes even lower, as they only need to retrain a portion of the graph. P-based methods are relatively controllable, and with patient training, they can achieve good unlearning efficacy. However, due to the disruption of the graph's topology, they may suffer from poor utility performance. Moreover, the prolonged retraining of multiple shards reduces the practicality.\\\vspace{-0.12in}
    \item \textbf{\model'Advantages}. Our method strikes a good balance between memory and time. The GPU memory required by our paradigm is comparable to that of P-based methods and retraining, and significantly smaller than that of IF-based methods. The processing time is also much shorter compared to P-based methods and retraining, while being comparable to IF on relatively large datasets. In our adversarial attack experiments, we achieve satisfactory pretrained \textbf{IE} in around 10-15 pretraining epochs, and when specific unlearning requests arise, fine-tuning can be completed in 3-5 epochs. These can be found in Fig.~\ref{fig:pretrain-finetune}.\\\vspace{-0.12in}
\end{itemize}

\subsection{Unlearning Case Study (RQ6)}
\begin{figure}[t]
    \centering
   \includegraphics[width=0.5\textwidth]{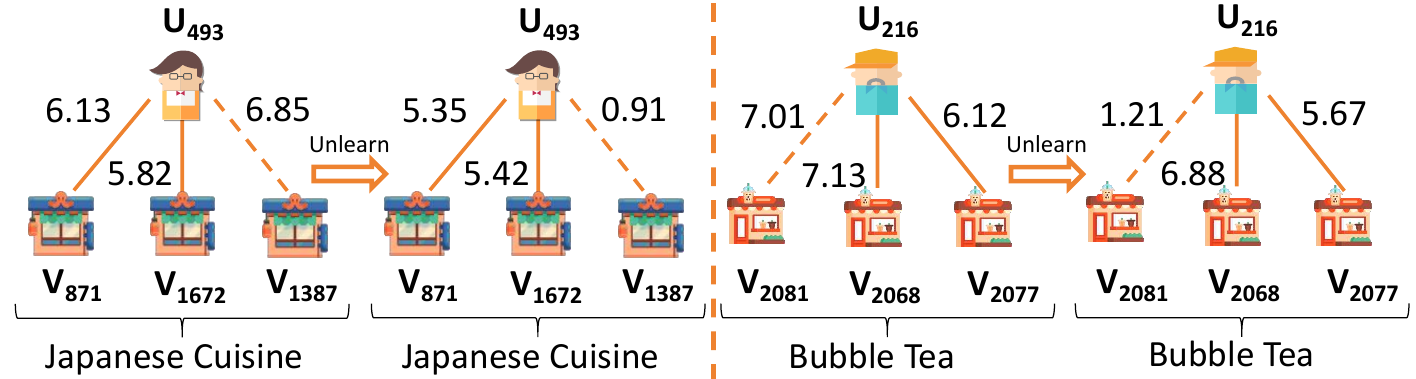}
    \vspace{-0.2in}    
    \caption{Case study on the unlearning effects for user-item interactions of similar categories based on Yelp2018 dataset.}
    \label{fig:case_study}
    \vspace{-0.1in}
\end{figure}

Here, we provide some examples of unlearning cases in Fig.~\ref{fig:case_study}. It can be observed that when a particular interaction is unlearned, its prediction score is significantly reduced (comparable to that of a negative interaction edge), and the prediction scores of items similar to it also decrease slightly. However, the scores for these similar items remain sufficiently high for them to be considered positive instances by the recommendation system and potentially recommended. When a user actively unlearns a significant number of similar items, the probability of recommended items from that category might also drop noticeably. This phenomenon aligns with our common sense: typically, when a user repeatedly removes interactions with the same type of items, it usually implies that they wish to no longer be recommended similar items. This demonstrates our method's strong generalization capability in unlearning tasks, along with its sophisticated modeling of complex dependencies and mutual influences between multiple unlearning requests.

\subsection{Unlearning Ratio Study}
Table~\ref{tab:un_ratio} summarizes the unlearning results of removing varying proportions of edges from the original dataset, evaluated through both model utility preservation and unlearning effectiveness metrics. In reality, unlearning requests typically only account for a small fraction of the dataset. From Tab.~\ref{tab:un_ratio}, it can be observed that as the proportion of unlearning increases, the predictive performance of the recommendation system slightly decreases in order to maintain a high level of unlearning efficacy. And our method maintains consistently high unlearning efficacy (MI-NG > 1 across all tests) when processing requests to remove varying proportions of user-item interactions. This demonstrates the robustness of our approach.

\begin{table}[t]
\caption{Utility and efficacy \textit{w.r.t.} different unlearning ratios.}
\label{tab:un_ratio}
\setlength{\tabcolsep}{0.7mm}
\vspace{-0.1in}
\resizebox{0.49\textwidth}{!}{
\begin{tabular}{c|ccccccc}
   \toprule[1.1pt]
   Unlearn Ratios & 0.10\% & 0.30\% & 0.50\% & 0.75\% & 1.00\% & 1.50\% &2.00\%  \\
   \midrule
   Recall &0.2564	&0.2543	&0.2504	&0.2462	&0.2411	&0.2321	&0.2302 \\
   NDCG  &0.1682 &0.1653 &0.1641 &0.1634 &0.1615 &0.1598 &0.1557 \\
   MI-BF &4.8497 &5.8521 &6.1612 &7.0485 &7.8845 &8.4836 &9.7556 \\
   MI-NG &1.1943 &1.1670 &1.0715 &1.0401 &1.0283 &1.0306 &1.0288\\
   \bottomrule[1.1pt]
\end{tabular}}
\vspace{-0.1in}
\end{table}

\section{Related Work}
\label{sec:relate}

\noindent \textbf{Partition-Based Approaches to Machine Unlearning}. Machine unlearning~\cite{cao2015towards, cong2022grapheditor,bourtoule2021machine} has garnered significant attention in recent years, with the goal of removing the influence of specific training data from model parameters while avoiding full retraining. Several approaches address these challenges across different domains. The \textit{SISA} concept~\cite{chen2022graph} partitions models into multiple shards and performs unlearning by retraining only the relevant shards. Similarly, \cite{li2024making} enhances collaborative filtering recommendation through optimized sequential training and hypergraph-based user clustering, enabling efficient recommendation unlearning. \\\vspace{-0.1in}

\noindent \textbf{Influence Function Methods for Unlearning}. Other efforts improve time efficiency by eliminating the retraining process entirely. Influence Function approaches represent a prominent direction, using mathematical methods to quickly estimate unlearning impact and directly modify model parameters with minimal computational overhead. For instance, \cite{guo2019certified} introduces certified data removal using influence functions, while \cite{izzo2021approximate} proposes approximate unlearning based on influence analysis for linear and logistic models. \cite{zhang2024recommendation} leverages influence functions for recommendation unlearning, achieving substantial speedup compared to retraining. A recent survey~\cite{li2024survey} systematizes design principles and taxonomies while highlighting challenges like dynamic unlearning and interpretability, underscoring the unique complexities of recommendation unlearning compared to traditional unlearning paradigms.

\section{Conclusion}
\label{sec:conclusoin}
In this work, we introduce \model: a novel pretraining paradigm for recommendation unlearning that is both model-agnostic and learnable. Our approach effectively addresses the complex challenges of unlearning in modern recommender systems, particularly excelling in advanced SSL-based architectures where traditional methods fall short. Through a strategic two-phase process of comprehensive pre-training followed by targeted fine-tuning with carefully designed optimization objectives, we demonstrate through extensive empirical evaluation that \model\ achieves a good balance. Our experiments across multiple benchmark datasets confirm that our method delivers exceptional unlearning performance while maintaining computational efficiency and memory utilization.

\clearpage

\bibliographystyle{ACM-Reference-Format}
\balance
\bibliography{refs}


\begin{thebibliography}{39}


\ifx \showCODEN    \undefined \def \showCODEN     #1{\unskip}     \fi
\ifx \showISBNx    \undefined \def \showISBNx     #1{\unskip}     \fi
\ifx \showISBNxiii \undefined \def \showISBNxiii  #1{\unskip}     \fi
\ifx \showISSN     \undefined \def \showISSN      #1{\unskip}     \fi
\ifx \showLCCN     \undefined \def \showLCCN      #1{\unskip}     \fi
\ifx \shownote     \undefined \def \shownote      #1{#1}          \fi
\ifx \showarticletitle \undefined \def \showarticletitle #1{#1}   \fi
\ifx \showURL      \undefined \def \showURL       {\relax}        \fi
\providecommand\bibfield[2]{#2}
\providecommand\bibinfo[2]{#2}
\providecommand\natexlab[1]{#1}
\providecommand\showeprint[2][]{arXiv:#2}

\bibitem[Act(2000)]%
        {act2000personal}
\bibfield{author}{\bibinfo{person}{Privacy Act}.} \bibinfo{year}{2000}\natexlab{}.
\newblock \showarticletitle{Personal information protection and electronic documents act}.
\newblock \bibinfo{journal}{\emph{Department of Justice, Canada. Full text available at http://laws. justice. gc. ca/en/P-8.6/text. html}} (\bibinfo{year}{2000}), \bibinfo{pages}{4356--4364}.
\newblock


\bibitem[Bourtoule et~al\mbox{.}(2021)]%
        {bourtoule2021machine}
\bibfield{author}{\bibinfo{person}{Lucas Bourtoule}, \bibinfo{person}{Varun Chandrasekaran}, \bibinfo{person}{Christopher~A Choquette-Choo}, \bibinfo{person}{Hengrui Jia}, \bibinfo{person}{Adelin Travers}, \bibinfo{person}{Baiwu Zhang}, \bibinfo{person}{David Lie}, {and} \bibinfo{person}{Nicolas Papernot}.} \bibinfo{year}{2021}\natexlab{}.
\newblock \showarticletitle{Machine unlearning}. In \bibinfo{booktitle}{\emph{2021 IEEE Symposium on Security and Privacy (SP)}}. IEEE, \bibinfo{pages}{141--159}.
\newblock


\bibitem[Cao and Yang(2015)]%
        {cao2015towards}
\bibfield{author}{\bibinfo{person}{Yinzhi Cao} {and} \bibinfo{person}{Junfeng Yang}.} \bibinfo{year}{2015}\natexlab{}.
\newblock \showarticletitle{Towards making systems forget with machine unlearning}. In \bibinfo{booktitle}{\emph{IEEE Symposium on Security and Privacy (S\&P)}}. IEEE, \bibinfo{pages}{463--480}.
\newblock


\bibitem[Chen et~al\mbox{.}(2022a)]%
        {chen2022recommendation}
\bibfield{author}{\bibinfo{person}{Chong Chen}, \bibinfo{person}{Fei Sun}, \bibinfo{person}{Min Zhang}, {and} \bibinfo{person}{Bolin Ding}.} \bibinfo{year}{2022}\natexlab{a}.
\newblock \showarticletitle{Recommendation unlearning}. In \bibinfo{booktitle}{\emph{Proceedings of the ACM Web Conference 2022}}. \bibinfo{pages}{2768--2777}.
\newblock


\bibitem[Chen et~al\mbox{.}(2025)]%
        {chen2025lightgnn}
\bibfield{author}{\bibinfo{person}{Guoxuan Chen}, \bibinfo{person}{Lianghao Xia}, {and} \bibinfo{person}{Chao Huang}.} \bibinfo{year}{2025}\natexlab{}.
\newblock \showarticletitle{LightGNN: Simple Graph Neural Network for Recommendation}. In \bibinfo{booktitle}{\emph{Proceedings of the Eighteenth ACM International Conference on Web Search and Data Mining}}. \bibinfo{pages}{549--558}.
\newblock
\href{https://doi.org/10.1145/3701551.3703536}{doi:\nolinkurl{10.1145/3701551.3703536}}


\bibitem[Chen et~al\mbox{.}(2020)]%
        {chen2020revisiting}
\bibfield{author}{\bibinfo{person}{Lei Chen}, \bibinfo{person}{Le Wu}, \bibinfo{person}{Richang Hong}, \bibinfo{person}{Kun Zhang}, {and} \bibinfo{person}{Meng Wang}.} \bibinfo{year}{2020}\natexlab{}.
\newblock \showarticletitle{Revisiting graph based collaborative filtering: A linear residual graph convolutional network approach}. In \bibinfo{booktitle}{\emph{AAAI Conference on Artificial Intelligence (AAAI)}}, Vol.~\bibinfo{volume}{34}. \bibinfo{pages}{27--34}.
\newblock


\bibitem[Chen et~al\mbox{.}(2022b)]%
        {chen2022graph}
\bibfield{author}{\bibinfo{person}{Min Chen}, \bibinfo{person}{Zhikun Zhang}, \bibinfo{person}{Tianhao Wang}, \bibinfo{person}{Michael Backes}, \bibinfo{person}{Mathias Humbert}, {and} \bibinfo{person}{Yang Zhang}.} \bibinfo{year}{2022}\natexlab{b}.
\newblock \showarticletitle{Graph unlearning}. In \bibinfo{booktitle}{\emph{ACM SIGSAC Conference on Computer and Communications Security (CCS)}}. \bibinfo{pages}{499--513}.
\newblock


\bibitem[Cong and Mahdavi(2022)]%
        {cong2022grapheditor}
\bibfield{author}{\bibinfo{person}{Weilin Cong} {and} \bibinfo{person}{Mehrdad Mahdavi}.} \bibinfo{year}{2022}\natexlab{}.
\newblock \showarticletitle{GraphEditor: An Efficient Graph Representation Learning and Unlearning Approach}.
\newblock  (\bibinfo{year}{2022}).
\newblock


\bibitem[Croitoru et~al\mbox{.}(2023)]%
        {croitoru2023diffusion}
\bibfield{author}{\bibinfo{person}{Florinel-Alin Croitoru}, \bibinfo{person}{Vlad Hondru}, \bibinfo{person}{Radu~Tudor Ionescu}, {and} \bibinfo{person}{Mubarak Shah}.} \bibinfo{year}{2023}\natexlab{}.
\newblock \showarticletitle{Diffusion models in vision: A survey}.
\newblock \bibinfo{journal}{\emph{IEEE Transactions on Pattern Analysis and Machine Intelligence (TPAMI)}} (\bibinfo{year}{2023}).
\newblock


\bibitem[Dukler et~al\mbox{.}(2023)]%
        {dukler2023safe}
\bibfield{author}{\bibinfo{person}{Yonatan Dukler}, \bibinfo{person}{Benjamin Bowman}, \bibinfo{person}{Alessandro Achille}, \bibinfo{person}{Aditya Golatkar}, \bibinfo{person}{Ashwin Swaminathan}, {and} \bibinfo{person}{Stefano Soatto}.} \bibinfo{year}{2023}\natexlab{}.
\newblock \showarticletitle{Safe: Machine unlearning with shard graphs}. In \bibinfo{booktitle}{\emph{IEEE/CVF International Conference on Computer Vision (ICCV)}}. \bibinfo{pages}{17108--17118}.
\newblock


\bibitem[Gao et~al\mbox{.}(2023)]%
        {gao2023survey}
\bibfield{author}{\bibinfo{person}{Chen Gao}, \bibinfo{person}{Yu Zheng}, \bibinfo{person}{Nian Li}, \bibinfo{person}{Yinfeng Li}, \bibinfo{person}{Yingrong Qin}, \bibinfo{person}{Jinghua Piao}, \bibinfo{person}{Yuhan Quan}, \bibinfo{person}{Jianxin Chang}, \bibinfo{person}{Depeng Jin}, \bibinfo{person}{Xiangnan He}, {et~al\mbox{.}}} \bibinfo{year}{2023}\natexlab{}.
\newblock \showarticletitle{A survey of graph neural networks for recommender systems: Challenges, methods, and directions}.
\newblock \bibinfo{journal}{\emph{ACM Transactions on Recommender Systems (TRS)}} \bibinfo{volume}{1}, \bibinfo{number}{1} (\bibinfo{year}{2023}), \bibinfo{pages}{1--51}.
\newblock


\bibitem[Guo et~al\mbox{.}(2019)]%
        {guo2019certified}
\bibfield{author}{\bibinfo{person}{Chuan Guo}, \bibinfo{person}{Tom Goldstein}, \bibinfo{person}{Awni Hannun}, {and} \bibinfo{person}{Laurens Van Der~Maaten}.} \bibinfo{year}{2019}\natexlab{}.
\newblock \showarticletitle{Certified data removal from machine learning models}.
\newblock \bibinfo{journal}{\emph{arXiv preprint arXiv:1911.03030}} (\bibinfo{year}{2019}).
\newblock


\bibitem[He et~al\mbox{.}(2020)]%
        {he2020lightgcn}
\bibfield{author}{\bibinfo{person}{Xiangnan He}, \bibinfo{person}{Kuan Deng}, \bibinfo{person}{Xiang Wang}, \bibinfo{person}{Yan Li}, \bibinfo{person}{Yongdong Zhang}, {and} \bibinfo{person}{Meng Wang}.} \bibinfo{year}{2020}\natexlab{}.
\newblock \showarticletitle{Lightgcn: Simplifying and powering graph convolution network for recommendation}. In \bibinfo{booktitle}{\emph{International ACM SIGIR conference on research and development in Information Retrieval (SIGIR)}}. \bibinfo{pages}{639--648}.
\newblock


\bibitem[Ho et~al\mbox{.}(2020)]%
        {ho2020denoising}
\bibfield{author}{\bibinfo{person}{Jonathan Ho}, \bibinfo{person}{Ajay Jain}, {and} \bibinfo{person}{Pieter Abbeel}.} \bibinfo{year}{2020}\natexlab{}.
\newblock \showarticletitle{Denoising diffusion probabilistic models}.
\newblock \bibinfo{journal}{\emph{Advances in Neural Informationf Processing Systems (NeurIPS)}}  \bibinfo{volume}{33} (\bibinfo{year}{2020}), \bibinfo{pages}{6840--6851}.
\newblock


\bibitem[Izzo et~al\mbox{.}(2021)]%
        {izzo2021approximate}
\bibfield{author}{\bibinfo{person}{Zachary Izzo}, \bibinfo{person}{Mary~Anne Smart}, \bibinfo{person}{Kamalika Chaudhuri}, {and} \bibinfo{person}{James Zou}.} \bibinfo{year}{2021}\natexlab{}.
\newblock \showarticletitle{Approximate data deletion from machine learning models}. In \bibinfo{booktitle}{\emph{International Conference on Artificial Intelligence and Statistics (AISTATS)}}. PMLR, \bibinfo{pages}{2008--2016}.
\newblock


\bibitem[Kwak et~al\mbox{.}(2017)]%
        {kwak2017let}
\bibfield{author}{\bibinfo{person}{Chanhee Kwak}, \bibinfo{person}{Junyeong Lee}, \bibinfo{person}{Kyuhong Park}, {and} \bibinfo{person}{Heeseok Lee}.} \bibinfo{year}{2017}\natexlab{}.
\newblock \showarticletitle{Let machines unlearn--machine unlearning and the right to be forgotten}.
\newblock  (\bibinfo{year}{2017}).
\newblock


\bibitem[Li et~al\mbox{.}(2024a)]%
        {li2024making}
\bibfield{author}{\bibinfo{person}{Yuyuan Li}, \bibinfo{person}{Chaochao Chen}, \bibinfo{person}{Xiaolin Zheng}, \bibinfo{person}{Junlin Liu}, {and} \bibinfo{person}{Jun Wang}.} \bibinfo{year}{2024}\natexlab{a}.
\newblock \showarticletitle{Making recommender systems forget: Learning and unlearning for erasable recommendation}.
\newblock \bibinfo{journal}{\emph{Knowledge-Based Systems}}  \bibinfo{volume}{283} (\bibinfo{year}{2024}), \bibinfo{pages}{111124}.
\newblock


\bibitem[Li et~al\mbox{.}(2024b)]%
        {li2024survey}
\bibfield{author}{\bibinfo{person}{Yuyuan Li}, \bibinfo{person}{Xiaohua Feng}, \bibinfo{person}{Chaochao Chen}, {and} \bibinfo{person}{Qiang Yang}.} \bibinfo{year}{2024}\natexlab{b}.
\newblock \showarticletitle{A Survey on Recommendation Unlearning: Fundamentals, Taxonomy, Evaluation, and Open Questions}.
\newblock \bibinfo{journal}{\emph{arXiv preprint arXiv:2412.12836}} (\bibinfo{year}{2024}).
\newblock


\bibitem[Li et~al\mbox{.}(2024c)]%
        {li2024recdiff}
\bibfield{author}{\bibinfo{person}{Zongwei Li}, \bibinfo{person}{Lianghao Xia}, {and} \bibinfo{person}{Chao Huang}.} \bibinfo{year}{2024}\natexlab{c}.
\newblock \showarticletitle{Recdiff: diffusion model for social recommendation}. In \bibinfo{booktitle}{\emph{International Conference on Information and Knowledge Management (CIKM)}}. \bibinfo{pages}{1346--1355}.
\newblock


\bibitem[Lin et~al\mbox{.}(2022)]%
        {lin2022improving}
\bibfield{author}{\bibinfo{person}{Zihan Lin}, \bibinfo{person}{Changxin Tian}, \bibinfo{person}{Yupeng Hou}, {and} \bibinfo{person}{Wayne~Xin Zhao}.} \bibinfo{year}{2022}\natexlab{}.
\newblock \showarticletitle{Improving graph collaborative filtering with neighborhood-enriched contrastive learning}. In \bibinfo{booktitle}{\emph{ACM Web Conference (WWW)}}. \bibinfo{pages}{2320--2329}.
\newblock


\bibitem[Olatunji et~al\mbox{.}(2021)]%
        {olatunji2021membership}
\bibfield{author}{\bibinfo{person}{Iyiola~E Olatunji}, \bibinfo{person}{Wolfgang Nejdl}, {and} \bibinfo{person}{Megha Khosla}.} \bibinfo{year}{2021}\natexlab{}.
\newblock \showarticletitle{Membership inference attack on graph neural networks}. In \bibinfo{booktitle}{\emph{2021 Third IEEE International Conference on Trust, Privacy and Security in Intelligent Systems and Applications (TPS-ISA)}}. IEEE, \bibinfo{pages}{11--20}.
\newblock


\bibitem[Pearlmutter(1994)]%
        {pearlmutter1994fast}
\bibfield{author}{\bibinfo{person}{Barak~A Pearlmutter}.} \bibinfo{year}{1994}\natexlab{}.
\newblock \showarticletitle{Fast exact multiplication by the Hessian}.
\newblock \bibinfo{journal}{\emph{Neural computation}} \bibinfo{volume}{6}, \bibinfo{number}{1} (\bibinfo{year}{1994}), \bibinfo{pages}{147--160}.
\newblock


\bibitem[Protection(2018)]%
        {protection2018general}
\bibfield{author}{\bibinfo{person}{Formerly~Data Protection}.} \bibinfo{year}{2018}\natexlab{}.
\newblock \showarticletitle{General data protection regulation (GDPR)}.
\newblock \bibinfo{journal}{\emph{Intersoft Consulting, Accessed in October}} \bibinfo{volume}{24}, \bibinfo{number}{1} (\bibinfo{year}{2018}).
\newblock


\bibitem[Rendle et~al\mbox{.}(2012)]%
        {rendle2012bpr}
\bibfield{author}{\bibinfo{person}{Steffen Rendle}, \bibinfo{person}{Christoph Freudenthaler}, \bibinfo{person}{Zeno Gantner}, {and} \bibinfo{person}{Lars Schmidt-Thieme}.} \bibinfo{year}{2012}\natexlab{}.
\newblock \showarticletitle{BPR: Bayesian personalized ranking from implicit feedback}.
\newblock \bibinfo{journal}{\emph{arXiv preprint arXiv:1205.2618}} (\bibinfo{year}{2012}).
\newblock


\bibitem[Sohl-Dickstein et~al\mbox{.}(2015)]%
        {sohl2015deep}
\bibfield{author}{\bibinfo{person}{Jascha Sohl-Dickstein}, \bibinfo{person}{Eric Weiss}, \bibinfo{person}{Niru Maheswaranathan}, {and} \bibinfo{person}{Surya Ganguli}.} \bibinfo{year}{2015}\natexlab{}.
\newblock \showarticletitle{Deep unsupervised learning using nonequilibrium thermodynamics}. In \bibinfo{booktitle}{\emph{International Conference on Machine Learning (ICML)}}. PMLR, \bibinfo{pages}{2256--2265}.
\newblock


\bibitem[Steihaug(1983)]%
        {steihaug1983conjugate}
\bibfield{author}{\bibinfo{person}{Trond Steihaug}.} \bibinfo{year}{1983}\natexlab{}.
\newblock \showarticletitle{The conjugate gradient method and trust regions in large scale optimization}.
\newblock \bibinfo{journal}{\emph{SIAM J. Numer. Anal.}} \bibinfo{volume}{20}, \bibinfo{number}{3} (\bibinfo{year}{1983}), \bibinfo{pages}{626--637}.
\newblock


\bibitem[Wang et~al\mbox{.}(2023)]%
        {wang2023diffusion}
\bibfield{author}{\bibinfo{person}{Wenjie Wang}, \bibinfo{person}{Yiyan Xu}, \bibinfo{person}{Fuli Feng}, \bibinfo{person}{Xinyu Lin}, \bibinfo{person}{Xiangnan He}, {and} \bibinfo{person}{Tat-Seng Chua}.} \bibinfo{year}{2023}\natexlab{}.
\newblock \showarticletitle{Diffusion Recommender Model}.
\newblock \bibinfo{journal}{\emph{International ACM SIGIR conference on research and development in Information Retrieval (SIGIR)}} (\bibinfo{year}{2023}).
\newblock


\bibitem[Wang et~al\mbox{.}(2019)]%
        {wang2019neural}
\bibfield{author}{\bibinfo{person}{Xiang Wang}, \bibinfo{person}{Xiangnan He}, \bibinfo{person}{Meng Wang}, \bibinfo{person}{Fuli Feng}, {and} \bibinfo{person}{Tat-Seng Chua}.} \bibinfo{year}{2019}\natexlab{}.
\newblock \showarticletitle{Neural graph collaborative filtering}. In \bibinfo{booktitle}{\emph{ACM SIGIR Conference on Research and Development in Information Retrieval (SIGIR)}}. \bibinfo{pages}{165--174}.
\newblock


\bibitem[Wu et~al\mbox{.}(2021)]%
        {wu2021self}
\bibfield{author}{\bibinfo{person}{Jiancan Wu}, \bibinfo{person}{Xiang Wang}, \bibinfo{person}{Fuli Feng}, \bibinfo{person}{Xiangnan He}, \bibinfo{person}{Liang Chen}, \bibinfo{person}{Jianxun Lian}, {and} \bibinfo{person}{Xing Xie}.} \bibinfo{year}{2021}\natexlab{}.
\newblock \showarticletitle{Self-supervised graph learning for recommendation}. In \bibinfo{booktitle}{\emph{International ACM SIGIR Conference on Research and Development in Information Retrieval (SIGIR)}}. \bibinfo{pages}{726--735}.
\newblock


\bibitem[Wu et~al\mbox{.}(2023b)]%
        {wu2023gif}
\bibfield{author}{\bibinfo{person}{Jiancan Wu}, \bibinfo{person}{Yi Yang}, \bibinfo{person}{Yuchun Qian}, \bibinfo{person}{Yongduo Sui}, \bibinfo{person}{Xiang Wang}, {and} \bibinfo{person}{Xiangnan He}.} \bibinfo{year}{2023}\natexlab{b}.
\newblock \showarticletitle{GIF: A General Graph Unlearning Strategy via Influence Function}. In \bibinfo{booktitle}{\emph{ACM Web Conference (WWW)}}. \bibinfo{pages}{651--661}.
\newblock


\bibitem[Wu et~al\mbox{.}(2023a)]%
        {wu2023certified}
\bibfield{author}{\bibinfo{person}{Kun Wu}, \bibinfo{person}{Jie Shen}, \bibinfo{person}{Yue Ning}, \bibinfo{person}{Ting Wang}, {and} \bibinfo{person}{Wendy~Hui Wang}.} \bibinfo{year}{2023}\natexlab{a}.
\newblock \showarticletitle{Certified edge unlearning for graph neural networks}. In \bibinfo{booktitle}{\emph{International Conference on Knowledge Discovery and Data Mining (KDD)}}. \bibinfo{pages}{2606--2617}.
\newblock


\bibitem[Wu et~al\mbox{.}(2022)]%
        {wu2022fast}
\bibfield{author}{\bibinfo{person}{Kun Wu}, \bibinfo{person}{Jie Shen}, \bibinfo{person}{Yue Ning}, {and} \bibinfo{person}{Wendy~Hui Wang}.} \bibinfo{year}{2022}\natexlab{}.
\newblock \showarticletitle{Fast Yet Effective Graph Unlearning through Influence Analysis}.
\newblock  (\bibinfo{year}{2022}).
\newblock


\bibitem[Wu et~al\mbox{.}(2020)]%
        {wu2020comprehensive}
\bibfield{author}{\bibinfo{person}{Zonghan Wu}, \bibinfo{person}{Shirui Pan}, \bibinfo{person}{Fengwen Chen}, \bibinfo{person}{Guodong Long}, \bibinfo{person}{Chengqi Zhang}, {and} \bibinfo{person}{S~Yu Philip}.} \bibinfo{year}{2020}\natexlab{}.
\newblock \showarticletitle{A Comprehensive Survey on Graph Neural Networks}.
\newblock \bibinfo{journal}{\emph{IEEE Transactions on Neural Networks and Learning Systems (TNNLS)}} \bibinfo{volume}{32}, \bibinfo{number}{1} (\bibinfo{year}{2020}), \bibinfo{pages}{4--24}.
\newblock


\bibitem[Xia et~al\mbox{.}(2023)]%
        {xia2023automated}
\bibfield{author}{\bibinfo{person}{Lianghao Xia}, \bibinfo{person}{Chao Huang}, \bibinfo{person}{Chunzhen Huang}, \bibinfo{person}{Kangyi Lin}, \bibinfo{person}{Tao Yu}, {and} \bibinfo{person}{Ben Kao}.} \bibinfo{year}{2023}\natexlab{}.
\newblock \showarticletitle{Automated Self-Supervised Learning for Recommendation}. In \bibinfo{booktitle}{\emph{The ACM Web Conference (WWW)}}. \bibinfo{pages}{992--1002}.
\newblock


\bibitem[Yu et~al\mbox{.}(2022)]%
        {yu2022graph}
\bibfield{author}{\bibinfo{person}{Junliang Yu}, \bibinfo{person}{Hongzhi Yin}, \bibinfo{person}{Xin Xia}, \bibinfo{person}{Tong Chen}, \bibinfo{person}{Lizhen Cui}, {and} \bibinfo{person}{Quoc Viet~Hung Nguyen}.} \bibinfo{year}{2022}\natexlab{}.
\newblock \showarticletitle{Are graph augmentations necessary? simple graph contrastive learning for recommendation}. In \bibinfo{booktitle}{\emph{International ACM SIGIR Conference on Research and Development in Information Retrieval (SIGIR)}}. \bibinfo{pages}{1294--1303}.
\newblock


\bibitem[Zhang et~al\mbox{.}(2022)]%
        {zhang2022incorporating}
\bibfield{author}{\bibinfo{person}{An Zhang}, \bibinfo{person}{Wenchang Ma}, \bibinfo{person}{Xiang Wang}, {and} \bibinfo{person}{Tat-Seng Chua}.} \bibinfo{year}{2022}\natexlab{}.
\newblock \showarticletitle{Incorporating bias-aware margins into contrastive loss for collaborative filtering}.
\newblock \bibinfo{journal}{\emph{Advances in Neural Information Processing Systems (NeurIPS)}}  \bibinfo{volume}{35} (\bibinfo{year}{2022}), \bibinfo{pages}{7866--7878}.
\newblock


\bibitem[Zhang et~al\mbox{.}(2019)]%
        {zhang2019deep}
\bibfield{author}{\bibinfo{person}{Shuai Zhang}, \bibinfo{person}{Lina Yao}, \bibinfo{person}{Aixin Sun}, {and} \bibinfo{person}{Yi Tay}.} \bibinfo{year}{2019}\natexlab{}.
\newblock \showarticletitle{Deep learning based recommender system: A survey and new perspectives}.
\newblock \bibinfo{journal}{\emph{ACM computing surveys (CSUR)}} \bibinfo{volume}{52}, \bibinfo{number}{1} (\bibinfo{year}{2019}), \bibinfo{pages}{1--38}.
\newblock


\bibitem[Zhang et~al\mbox{.}(2024)]%
        {zhang2024recommendation}
\bibfield{author}{\bibinfo{person}{Yang Zhang}, \bibinfo{person}{Zhiyu Hu}, \bibinfo{person}{Yimeng Bai}, \bibinfo{person}{Jiancan Wu}, \bibinfo{person}{Qifan Wang}, {and} \bibinfo{person}{Fuli Feng}.} \bibinfo{year}{2024}\natexlab{}.
\newblock \showarticletitle{Recommendation unlearning via influence function}.
\newblock \bibinfo{journal}{\emph{ACM Transactions on Recommender Systems}} \bibinfo{volume}{3}, \bibinfo{number}{2} (\bibinfo{year}{2024}), \bibinfo{pages}{1--23}.
\newblock


\bibitem[Zhou et~al\mbox{.}(2020)]%
        {zhou2020graph}
\bibfield{author}{\bibinfo{person}{Jie Zhou}, \bibinfo{person}{Ganqu Cui}, \bibinfo{person}{Shengding Hu}, \bibinfo{person}{Zhengyan Zhang}, \bibinfo{person}{Cheng Yang}, \bibinfo{person}{Zhiyuan Liu}, \bibinfo{person}{Lifeng Wang}, \bibinfo{person}{Changcheng Li}, {and} \bibinfo{person}{Maosong Sun}.} \bibinfo{year}{2020}\natexlab{}.
\newblock \showarticletitle{Graph neural networks: A review of methods and applications}.
\newblock \bibinfo{journal}{\emph{AI open}}  \bibinfo{volume}{1} (\bibinfo{year}{2020}), \bibinfo{pages}{57--81}.
\newblock


\end{thebibliography}

\end{document}